\documentclass[twocolumn]{aastex63}

\submitjournal{ApJ}

\usepackage{graphicx}
\usepackage{rotating,booktabs}
\usepackage[verbose]{placeins}
\usepackage[english]{babel}
\usepackage{chngpage}
\usepackage{amssymb}
\usepackage{enumitem}
\usepackage{txfonts}

\usepackage{natbib}

\newcommand{\gsim}{\raisebox{-0.13cm}{~\shortstack{$>$ \\[-0.07cm]$\sim$}}~}
\newcommand{\lsim}{\raisebox{-0.13cm}{~\shortstack{$<$ \\[-0.07cm]$\sim$}}~}

\shortauthors{Travascio, A.; Fabbiano, G.; Paggi, A., Elvis, M. et al.}
\graphicspath{{./}{figures/}}

\begin{document}

\title{\large AGN-host interaction in IC 5063. I. Large-scale X-ray morphology and spectral analysis}

\correspondingauthor{Travascio, A.}
\email{andrea.travascio@inaf.it}

\author[0000-0002-8863-888X]{Travascio, A.}
\affiliation{INAF-Osservatorio Astronomico di Trieste, via G.B. Tiepolo 11, 34143 Trieste, Italy}

\author{Fabbiano, G.}
\affiliation{Center for Astrophysics $\mid$ Harvard \& Smithsonian, 60 Garden Street, Cambridge, MA 02138, USA}

\author{Paggi, A.}
\affiliation{Dipartimento di Fisica, Universita' degli Studi di Torino, via Pietro Giuria 1, I-10125 Torino, Italy}
\affiliation{Istituto Nazionale di Fisica Nucleare, Sezione di Torino, via Pietro Giuria 1, I-10125 Torino, Italy}
\affiliation{INAF- Osservatorio Astrofisico di Torino, via Osservatorio 20, 10025 Pino Torinese, Italy}

\author{Elvis, M.}
\affiliation{Center for Astrophysics $\mid$ Harvard \& Smithsonian, 60 Garden Street, Cambridge, MA 02138, USA}

\author{Maksym, W. P.}
\affiliation{Center for Astrophysics $\mid$ Harvard \& Smithsonian, 60 Garden Street, Cambridge, MA 02138, USA}

\author{Morganti, R.}
\affiliation{ASTRON, Netherlands Institute for Radio Astronomy, Oude Hoogeveensedijk 4, 7991 PD, Dwingleoo, the Netherlands}
\affiliation{Kapteyn Astronomical Institute, University of Groningen, PO Box 800, 9700 AV, Groningen, the Netherlands}

\author{Oosterloo, T.}
\affiliation{ASTRON, Netherlands Institute for Radio Astronomy, Oude Hoogeveensedijk 4, 7991 PD, Dwingleoo, the Netherlands}
\affiliation{Kapteyn Astronomical Institute, University of Groningen, PO Box 800, 9700 AV, Groningen, the Netherlands}

\author{Fiore, F.}
\affiliation{INAF-Osservatorio Astronomico di Trieste, via G.B. Tiepolo 11, 34143 Trieste, Italy}

\nocollaboration{8}



\begin{abstract}
We report the analysis of the deep ($\sim$270~ks) X-ray $Chandra$ data of one of the most radio-loud, Seyfert 2 galaxies in the nearby Universe (z=0.01135), IC 5063. The alignment of the radio structure with the galactic disk and ionized bi-cone, enables us to study the effects of both radio jet and nuclear irradiation on the inter-stellar medium (ISM).
The nuclear and bi-cone spectra suggest a low photoionization phase mixed with a more ionized or thermal gas component, while the cross-cone spectrum is dominated by shocked and collisionally ionized gas emission. The clumpy morphology of the soft ($<$3~keV) X-ray emission along the jet trails, and the large ($\simeq$2.4~kpc) filamentary structure perpendicular to the radio jets at softer energies ($<$1.5~keV), suggest a large contribution of the jet-ISM interaction to the circumnuclear gas emission. 
The hard X-ray continuum ($>$3~keV) and the Fe K$\alpha$ 6.4~keV emission are both extended to kpc size along the bi-cone direction, suggesting an interaction of nuclear photons with dense clouds in the galaxy disk, as observed in other Compton Thick (CT) active nuclei. The north-west cone spectrum also exhibits an Fe XXV emission line, which appears spatially extended and spatially correlated with the most intense radio hot-spot, suggesting jet-ISM interaction.

\end{abstract}



\section{Introduction}\label{sec:intro}
Accreting supermassive black holes (SMBHs) in galactic nuclei release a large amount of energy in the form of radiation, winds, and radio jets, which interact with the interstellar medium (ISM) affecting the host galaxy evolution \citep[e.g.][]{Silk98,DiMatteo05}. The process linking the AGN energy and the surrounding environment is called AGN feedback. 

In very luminous quasars, radiative-driven outflows (kinetic-mode feedback) are thought to play a crucial role in the SMBH host galaxy co-evolution \citep{Ferrarese05}, by transferring less than 10$\%$ of the energy and momentum to the ISM \citep{Fiore17}. The most extreme example of AGN feedback occurs in very massive galaxies in cluster cores through radio jets (jet-mode feedback), which push away the hot X-ray emitting gas, producing cavities and shock fronts in the intra-cluster medium, and regulate temperature and entropy in there \citep{Gitti12,Russell13,Gupta20}.

However, moderate radio jets are also observed in local Seyfert galaxies. These jets are less powerful and extended than those in powerful radio galaxies, and interact with the gas on galactic scales, the ISM \citep{Morganti99,Thean00}.
While the complex nature of the interaction is still under study, there is evidence that radio jets can produce galactic winds which, in turn, interact with dense multiphase gas \citep[][and references therein]{Rosario08,Riffel14}. 
Both neutral and ionized outflows have been observed in radio galaxies \citep{Morganti03,Emonts05,Holt08,Lehnert11,Dasyra12,Combes13}. Studying the origin and the impact of these processes is fundamental for a total understanding of feedback \citep{Mukherjee16}.

In this paper we analyse the X-ray $Chandra$ data of the nearby (z=0.01135\footnote{From the NASA/IPAC Extragalactic Database.}) elliptical, type II \citep{Inglis93} galaxy, IC 5063. It is one of the most radio-loud \citep[$\rm P_{1.4~GHz} = 6.3 \times 10^{23} W~Hz^{-1}$][]{Morganti98} Seyfert 2 galaxies in the local Universe. It shows a ionizing radiation field with a ``X'' morphology \citep{Colina91} and a complex system of dust-lanes, likely due to merger remnants \citep{Morganti98,Oosterloo17}.
High-resolution radio data \citep[ATCA at 8 GHz][]{Morganti98} reveal a triple radio structure (1.3 kpc) along the PA of $\sim$295$^o$ with a total flux density of 230 mJy, consisting of a nuclear blob and two hot-spots, i.e. termination points where the jets collide with the gas; most of the flux (195 mJy) is emitted from  the northwest radio hot-spot. 

IC 5063 is a perfect laboratory to explore with $Chandra$ both the extended X-ray emission \citep[e.g.][]{Levenson06}, which is not diluted by the nuclear continuum, and the physical processes involved in the jet-ISM interaction. Indeed, it is a rare system because the radio jets lie to the same plane of the HI galactic disk \citep[PA$\sim$300$^o$][]{Danziger81,Morganti98} allowing a full interaction between the mechanical energy released by an AGN and the host galaxy.

One of the effects observed from the coupling between jet and ISM gas in IC 5063 is large-scale outflows. \cite{Morganti98} found the first case of an AGN-driven massive outflow of neutral HI gas in IC 5063 close to the bright NW radio lobe. 
Outflowing components are detected in warm ionized gas \citep{Morganti07,Dasyra15,Venturi21}, atomic \citep{Oosterloo00}, and warm and cold molecular gas \citep{Tadhunter14,Morganti15,Dasyra16,Oosterloo17}, with velocity dispersions $\sim$700$~\rm km~s^{-1}$. 
Other interesting features are observed perpendicular to the radio jet, including an extension of the radio continuum at 1.4 GHz \cite{Morganti98}, a giant low-ionization ([NII] and [SII]) loop by \cite{Maksym20}, and high velocity dispersion of H$\alpha$ and [OIII] emission lines by \cite{Venturi21}. 
These features suggest a lateral outflow, in agreement with the simulation of \cite{Mukherjee18} of IC 5063.

The analysis of the X-ray observations of IC 5063 allows us to study both the innermost emission of the AGN and the effects of the radio jet on the hot circumnuclear gas. The depth of our $Chandra$ data enables us to perform spatially resolved analyses, constraining the dominant processes and allowing us to investigate the feedback scenario in this specific case.

This is the first paper based on our study of the properties of the X-ray hot gas in IC 5063 with deep (272 ks) $Chandra$ observations (PI G. Fabbiano). IC 5063 has been already observed in X-rays with ASCA \citep{Vignali97}, ROSAT \citep{Pfeffermann87} and Suzaku broadband plus Swift BOSS \citep{Tazaki11}. 
However, it is only with the high spatial resolution of $Chandra$ that we can separately study the spectral properties of the point-like nuclear source and of the surrounding extended X-ray emission. Here we also investigate the spectral dependence of the morphology of this extended emission, following the procedures outlines in previous studies of AGNs with $Chandra$ \cite[e.g.][]{Wang11a,Paggi12,Feruglio13,Fabbiano18a,Maksym19,Jones20}.

This paper is organized as follows.
In Sect.~\ref{sec:DataReduction} we describe the X-ray $Chandra$ data, and we show the procedure of reduction and alignment applied on these observations, to optimise the 1/8 sub-pixel analysis. 
We report the results of the spatial analysis of the X-ray emission at different energy bands in Sect.~\ref{sect:imradprof}. 
We then perform the spectral analysis both for the nuclear and for the extended emission in Appendix.~\ref{sec:procedure1} and Appendix.~\ref{sec:procedure2}, respectively. 
Finally, we discuss our results in Sect.~\ref{sec:discussion} and summarize our findings and conclusions in Sect.~\ref{sec:conclusion}.


Throughout this paper we adopt cosmological parameters $H_0=67.7$, $\Omega _{\Lambda, 0}= 0.69$ and $\Omega _{m, 0}=0.31$ \citep{Planck18}. At the redshift of IC 5063 (luminosity distance $\rm D_L =$50.7 Mpc) the physical scale is $\sim$ 240 pc $\rm arcsec^{-1}$.



\section{Data reduction and Analysis} \label{sec:DataReduction}

In this paper we analyse six $Chandra$ ACIS-S (Advanced CCD Imaging Spectrometer-Spectral component) subarray mode observations of IC~5063 (Table~\ref{table:ObsID}) obtained in 2018/2019 with a cumulative exposure time of 238 ks (P.I. Fabbiano). These observations are combined with an archival $Chandra$ ACIS-S full-array mode observation (ObsID: 7878; P.I. D. Evans) obtained in 2007, reaching a cumulative effective exposure time of 272 ks. 

The data were reprocessed following the standard pipeline with the $Chandra$ Interactive Analysis of Observations \citep[\texttt{CIAO 4.12}][]{Fruscione06} and the $Chandra$ Calibration Data Base \citep[\texttt{CALDB 4.9.0}][]{Graessle07}.
We removed background flares exceeding 3$\sigma$ from the light curve of the single observations using the \textit{lc\_sigma\_clip} task\footnote{$https://cxc.harvard.edu/ciao/ahelp/lc\_sigma\_clip.html$}, based on an iterative sigma-clipping algorithm, thus reaching a final cumulative exposure time of $\sim$268 ks.

\setlength{\tabcolsep}{2pt}
\begin{table}[!t] 
\begin{adjustwidth}{-1.8em}{}
\caption{$Chandra$ observation log.}
\label{table:ObsID}
\begin{tabular}{ccccc}
\hline
\hline
ObsID                 &  Instrument   &     $\rm T_{exp}~[ks]$ & PI & Date    \\[3pt]
\hline    
7878  & ACIS & 34.1 & D. Evans & 2007-06-15/16 \\
21467  & ACIS & 26.9 & G. Fabbiano & 2018-12-11 \\
21999  & ACIS & 34.1 & G. Fabbiano & 2018-12-12/13 \\
22000  & ACIS & 15.6 & G. Fabbiano & 2018-12-13/14 \\
22001  & ACIS & 29.3 & G. Fabbiano & 2018-12-15 \\
22002  & ACIS & 43.9 & G. Fabbiano & 2018-12-16 \\
21466  & ACIS & 87.7 & G. Fabbiano & 2019-07-23/24 \\
\hline
\hline
\end{tabular}
\end{adjustwidth}
\end{table}

\subsection{Merging} \label{sec:merging}

\begin{figure*}
   \begin{center}
   \includegraphics[height=0.365\textheight,angle=0]{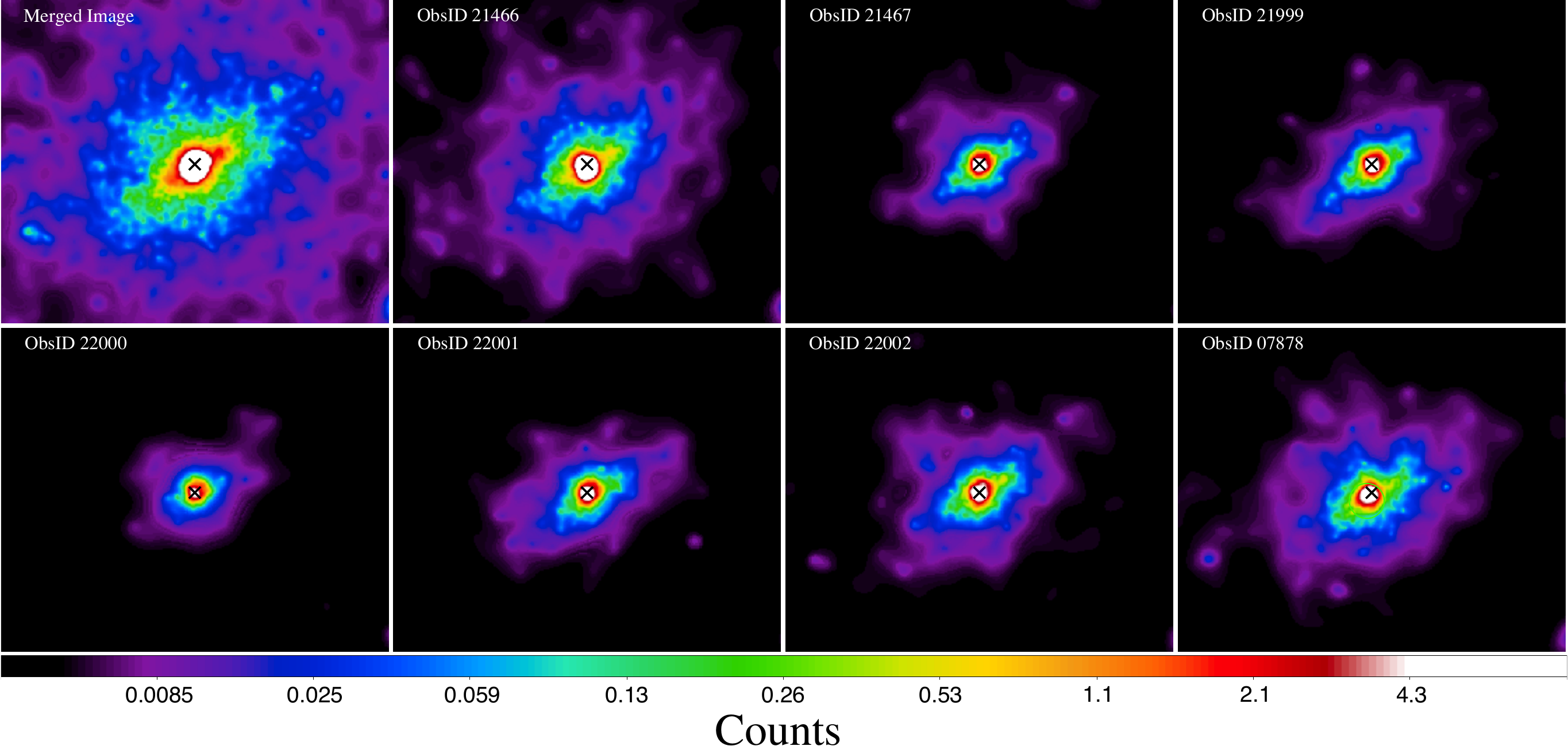}
   \caption{Adaptively smoothed (using \texttt{dmimgadapt}; see text for details), 1/8 sub-pixel images of the all observations plus the final merged image (top-left first), which is obtained by aligning all seven images. N is up and E to the left. In each image we report a black cross indicating the counts peak position used to align the sub-array mode observations, and we identify the data from which the image was derived. The only full-array mode observation, i.e. ObsID 7878, is aligned by matching the X-ray point sources in the field with respect to the deepest observation, i.e. the ObsID 21466 (see text for details).}
   \label{image:IC5063}
   \end{center}
\end{figure*}

\begin{figure*}
   \begin{center}
   \includegraphics[height=0.36\textheight,angle=0]{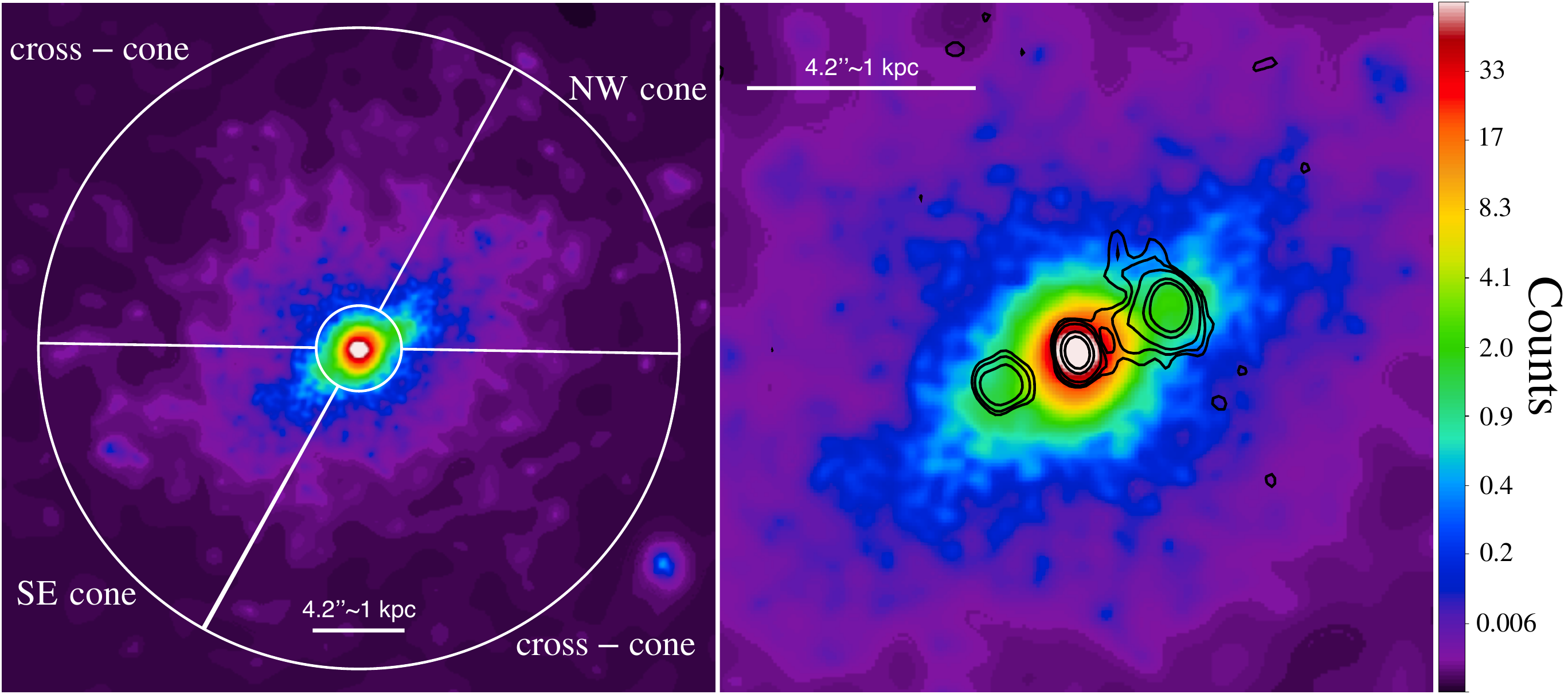}
   \caption{Left panel reports the adaptively smoothed (similarly to the images in Fig.~\ref{image:IC5063}), 1/8 pixel image of the emission at the 0.3-7.0 keV band. White full lines divide the nucleus of IC 5063 from the extended ($>$2$''$) region which, in turn, is split in four regions with opening angles of $\approx$60 and 120$^{\circ}$, to separate the bi-cone from the cross-cone area. Specifically, the bi-cone regions are enclosed in North-West (NW) and South-East (SE) sectors and limited at the position angles 269$^{\circ}$-331$^{\circ}$ and 89$^{\circ}$-151$^{\circ}$, respectively. In the right panel, we show the zoom-in of the IC 5063 nucleus. Black contours represent the radio emission observed with ATCA at 17 GHz and levels 3,5,10,50,100$\sigma$ \citep[$\rm \sigma \approx 0.11~ mJy~beam^{-1}$;][]{Morganti07}.}
   \label{image:regions} 
   \end{center}
\end{figure*}

The $Chandra$ observations are merged using the CIAO tool \texttt{reproject\_obs}\footnote{$https://cxc.cfa.harvard.edu/ciao/ahelp/reproject\_obs.html$}, adopting the deepest observation ObsID 21466 as reference frame. 
To produce the most accurate merged data set to enable sub-pixel analysis with image pixel size 1/8 ACIS pixel (0.492$''$), we explored different ways of aligning the data from each observation.

\subsubsection{Aligning off-nuclear point sources}\label{merg1}

We used off-nuclear point-like sources in the field of view, $<$2$''$ away from nuclear point source, to get a first alignment with the \texttt{reproject\_aspect} tool (hereinafter off-nuclear alignment method). These sources were detected with the CIAO \texttt{wavdetect} tool, adopting $\sqrt{2}$-series\footnote{i.e. "1.0 1.414 2.0 2.828 4.0 5.657 8.0 11.314 16.0", which are used to get a more extensive run. See $\rm http://www.cr.scphys.kyoto\-u.ac.jp/old\_html/local/chandra/detect.pdf$ as reference} scales as wavelet parameter and a threshold significance of $10^{-6}$ false detections per pixel.

By comparing the offsets of the point source centroids in the various observations relative to the final merged image in the 0.3-7.0 keV energy band, we find that this method yields a position accuracy of $\sim$0.5 instrumental ACIS pixel.

\subsubsection{Aligning the nuclear peak positions}\label{merg2}

Another method we use to obtain a merged data is to align the counts peak positions of the nuclear source in each observation (hereinafter nuclear alignment method). 

Given the high count rate of the bright nuclear source, the position of the counts peak could be affected by pileup\footnote{$https://cxc.cfa.harvard.edu/ciao/ahelp/acis\_pileup.html$}. Except for ObsID 7878, the ACIS observations were performed in subarray mode to minimize pileup. However, using the CIAO \texttt{pileup\_map} tool, we find that the pileup fraction reaches 16$\%$ and 33$\%$ per pixel in the 0.3-7.0 keV subarray and full-array observations, respectively.

To evaluate the effect of pileup in estimating the correct counts peak positions of the nuclear source, we simulated the Point-Spread Function (PSF) with and without pileup in the 0.3-7.0 keV energy band, using the $Chandra$ Ray Tracer\footnote{$https://cxc.cfa.harvard.edu/ciao/PSFs/chart2/$} \citep[\texttt{ChaRT}][]{Carter03} and \texttt{MARX 5.5.0}\footnote{$https://space.mit.edu/cxc/marx/$} tools. We then fitted these PSF models with a 2D Gaussian function, and measured the average spatial offsets between the Gaussian peak position of the PSFs simulated with and without pileup.
We find that the average spatial offsets in the subarray observations is $\sim \rm 3~mas$, $< 1/10$ of the native ACIS pixel. Instead, in the full-array observation (ObsID 7878), which accounts for 13$\%$ of the total exposure, the offset is $\sim$0.77 native pixel. Therefore, the pileup contribution to the counts peak position in the sub-pixel, full-array image is not negligible. 

Based on this result, we used the nuclear alignment method to merge the subarray mode observations only. The image of each observation was binned at 1/8 of the native pixel size using a sub-pixel event repositioning procedure\footnote{$https://cxc.harvard.edu/ciao4.7/why/acissubpix.html$}, and smoothed with the CIAO \texttt{aconvolve}\footnote{$https://cxc.cfa.harvard.edu/ciao/ahelp/aconvolve.html$} tool using a Gaussian kernel of 3 image pixels. We derived the counts peak positions in these images at the 6.1-6.6 keV energy band, which is expected to be the most point-like.
After re-projecting each image with \texttt{wcs\_update}\footnote{$https://cxc.cfa.harvard.edu/ciao/ahelp/wcs\_update.html$}, we combined them by producing a nuclear-aligned merged image. 

To select the most accurate merging method we compared the Full-Width at Half Maximum (FWHM) of the Gaussian components modelling the nuclear source in the merged images in the 6.1-6.6 keV energy band. To fit the nuclear source we used a rotating elliptical Gaussian model, i.e. the function \texttt{gauss2d}\footnote{$https://cxc.cfa.harvard.edu/sherpa/ahelp/gauss2d.html$} in \texttt{Sherpa}\footnote{$https://cxc.harvard.edu/sherpa/$}, and we extracted the FWHM of the best-fit 2D Gaussian models along the major axis.
We find that the FWHM in the 6.1-6.6 keV merged image from the nuclear alignment method is $1.414 \pm 0.026$ native pixels, $\sim$1.2 times the pre-flight PSF-size in the same energy band (FWHM$\simeq 1.213 _{-0.028}^{+0.022}$ native pixel). A slightly larger FWHM ($1.453 \pm 0.024$ native pixels) is obtained by fitting the nuclear source in the off-nuclear point-sources merged image at 6.1-6.6 keV. Although these values are consistent at 1$\sigma$, according to this comparison the method of aligning the peaks of the nuclear counts produces a most accurate merged image.

\subsubsection{Merging subarray and full-array mode observations}\label{merg3}

To obtain the final merged image, we combined the nuclear-aligned merged image and the full-array observation (7878) by aligning their off-nuclear point-like sources (Sect.~\ref{merg1}). 
From this image, we estimated a FWHM$= 1.371_{-0.024}^{+0.022}$ native pixel to the Gaussian model, lower than the one derived from the nuclear-aligned merged image, but consistent at $\sim$1.3$\sigma$.

Fig.~\ref{image:IC5063} shows both the re-projected images of each observation and the merged image obtained through the best alignment method. These are 1/8 sub-pixel images, adaptively smoothed with the \texttt{dmimgadapt}\footnote{$https://cxc.cfa.harvard.edu/ciao/ahelp/dmimgadapt.html$} tool, adopting a minimum and maximum smoothing logarithmic scales of 2 and 15 native pixel, a minimum number of 4 counts under the kernel and 15 scales to use, at the energy band 0.3-7.0 keV. The black crosses mark the peak positions (at R.A.=20:52:02.311,  decl.=-57:04:07.623) used as reference for the alignment of the subarray mode observations. 

\begin{figure*}[t]
   \begin{center}
   \includegraphics[height=0.63\textheight,angle=0]{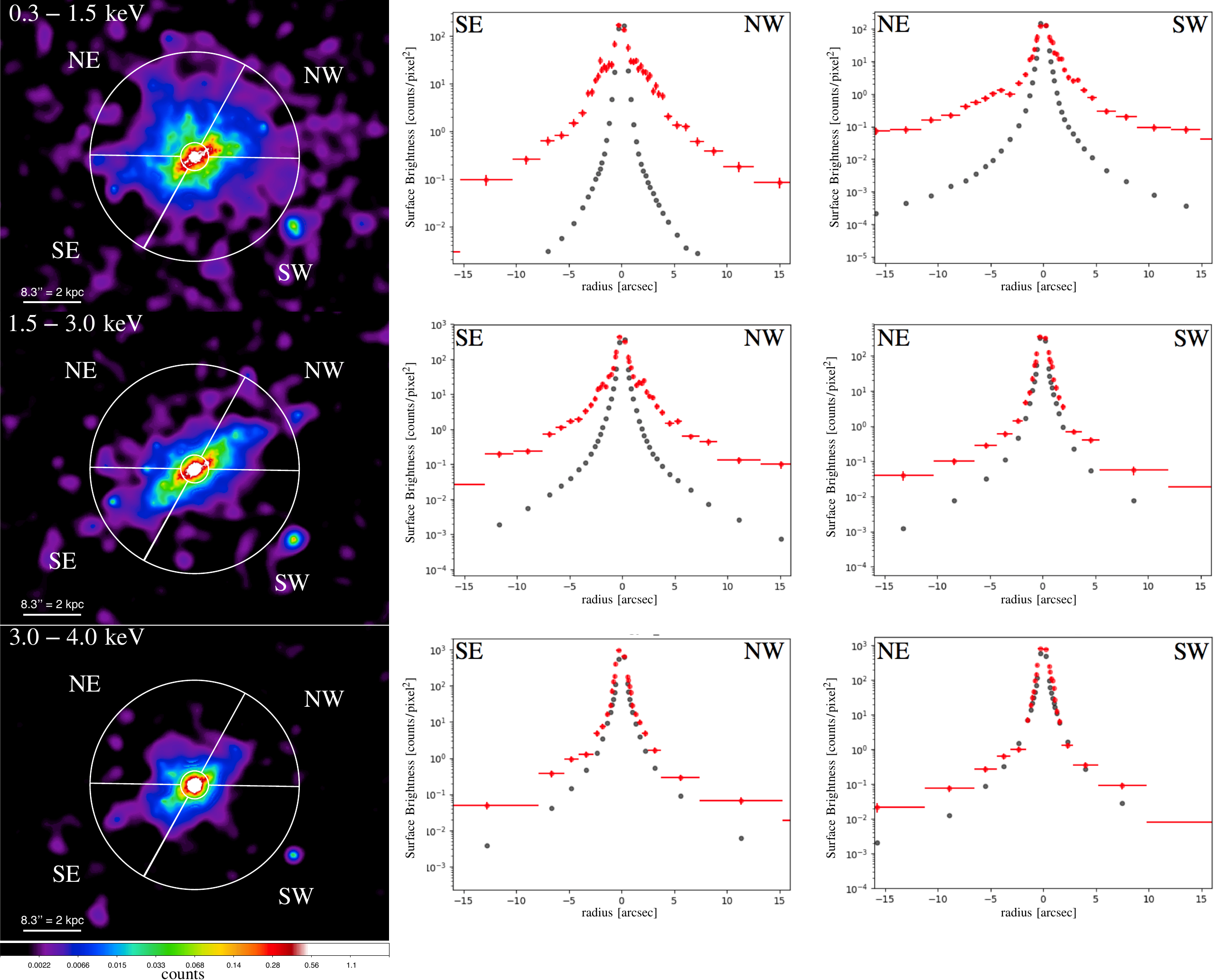}
   \caption{Images (left column) and surface brightness radial profiles in the bi-cone (central column) and cross-cone (right column) sectors, of IC 5063 in different energy bands ($top~panels$: 0.3-1.5~keV, $middle~panels$: 1.5-3.0~keV, $bottom~panels$: 3.0-4.0~keV). The images are re-binned at 1/8 of native pixel and adaptively smoothed with \texttt{dmimgadapt} (setting the following parameters: min=1, max=15, num=12, radscale=log, counts=5). The radial profiles are background-subtracted and estimated from the 1/8 sub-pixel images (red) and from the normalized PSF-model images (black). Uncertainties are 1$\sigma$ and the bin size was chosen to contain a minimum of 25 counts.}
   \label{image:rp1}
   \end{center}
\end{figure*}
\begin{figure*}[t]
   \begin{center}
   \includegraphics[height=0.8\textheight,angle=0]{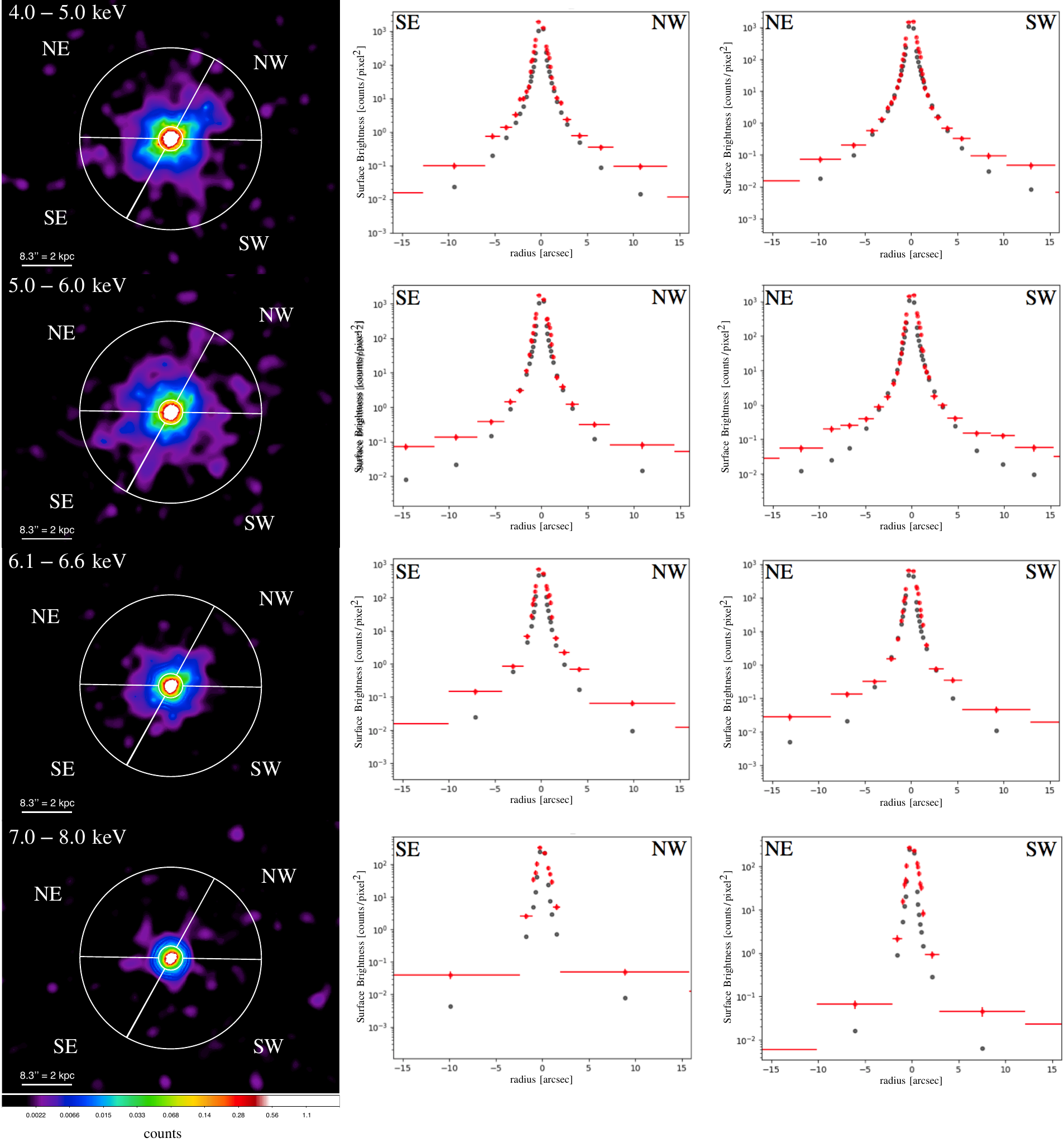}
   \caption{$Top~panels$: 4.0-5.0~keV, $top/middle~panels$: 5.0-6.0~keV, $bottom/middle~panels$: 6.1-6.6~keV, $bottom~panels$: 7.0-8.0~keV. See caption in the Fig.~\ref{image:rp1} for details.}
   \label{image:rp2}
   \end{center}
\end{figure*}

\section{Images and Radial Profiles}\label{sect:imradprof}

The final merged image in Fig.~\ref{image:regions} shows a prominent point-like nuclear source and fainter X-ray emission extended $\rm \sim 12''$ ($\sim$3 kpc) from the nucleus, in the South-East to North-West direction (P.A.$\sim$295$^{\circ}$-300$^{\circ}$). This is also the direction of the ionization cone \citep{Colina91} and the radio jets \citep{Morganti98}. 
Based on this morphology, we sliced the image in bi-cone and cross-cone sectors, to investigate separately the circum-nuclear gas morphology and extent.
To optimally determine the angles of these cones, we produced a surface brightness azimuthal profile of the 0.3-7.0 keV merged image within the 1$''$-10$''$ annulus centered on the nucleus.
We fitted this profile with a constant plus two Gaussian components, with peaks 180 degrees apart and the same width ($\sigma_G$). We defined the bisector and the opening angles of the bi-cone sectors as the Gaussian peaks and $\rm \pm 3 \sigma_G$ widths of the Gaussian profiles, respectively. 
The resulting bi-cone areas are enclosed within P.A.$\sim$89$^{\circ}$ and 151$^{\circ}$, with opening angle of $\sim$62$^{\circ}$. 

We produced 1/8 sub-pixel images in seven energy bands (left column of Figs.~\ref{image:rp1},\ref{image:rp2}). Each image is adaptively smoothed with the \textit{dmimgadapt} tool by adopting a minimum (maximum) smoothing logarithmic scale of 1 (15), a minimum of 5 counts under kernel and 30 iterations.

To estimate the extent of the emission in each energy band, we followed the procedure of \cite{Fabbiano18a} and \cite{Jones20}. The rightmost columns in Figs.~\ref{image:rp1} and \ref{image:rp2} show a comparison between the radial surface brightness profiles of the data (red points) and the PSF profiles (black points) in $\rm counts~bin^{-1}$ units, in the different energy bands. 
The radial surface brightness profiles were extracted from the 1/8 sub-pixel images, with bin sizes varying to contain a minimum of 25 counts. We subtracted the background counts estimated from the 11$''$ circle region centered at R.A.=20:52:05.9100 and decl.=-57:03:43.750, as well as the contribution of off-nuclear point sources, at R.A.= 20:52:00.5582, decl.=-57:04:17.784 and R.A.= 20:51:57.7932, decl.=57:04:06.178.
The PSF radial profiles were extracted from PSF images, which are obtained as the average of 500 simulated pileup-corrected PSF models, and normalized to the counts within 1$''$ radius centered in the nuclear source.

\section{SPECTRAL ANALYSIS AND RESULTS}\label{Spec}

Based on the bi-cone and cross-cone areas obtained in Sect.~\ref{sect:imradprof}, we selected four regions from which to extract the spectral data (1) a 2$''$ radius circle, enclosing approximately the 90$\%$ of the nuclear point source at the effective energy of 1 keV, and three sectors of the 2$''$-15$''$ annulus we name (2) North-West (NW) cone, (3) South-East (SE) cone and (4) cross-cone (shown in Fig.~\ref{image:regions}).

We performed separated spectral analyses of the NW and SE cones, because previous works on IC 5063 shows that the two cone regions host multi-phase gas with different properties (i.e. density, kinematics; e.g. \citealt{Dasyra16,Morganti07}). We find that the two bi-cone spectra are somewhat different. The spectra extracted from the two sectors in the cross-cone region do not show significant differences.

The spectral data were extracted with the CIAO \texttt{specextract}\footnote{$https://cxc.cfa.harvard.edu/ciao/ahelp/specextract.html$} script.
All the spectra were background subtracted from a circular region of 15$''$ radius, located 50$''$ away from the nuclear source, free of X-ray point-like sources.
The spectra were binned to have a minimum of 20 counts per bin, and fitted with \texttt{Sherpa}. 
For each spectrum, we included a constant Galactic absorption model with weighted average column density $\rm N_H = 5.77 \times 10^{20} cm^{-2}$ derived with the NASA HEASARC tool\footnote{$https://heasarc.gsfc.nasa.gov$}. This Galactic hydrogen column density was computed within a cone radius of 0.1 degree. 
The best-fit parameters errors are reported at 1$\sigma$ of confidence level.

For each spectrum, we first performed a fit with a phenomenological model in order to characterize the spectral shape and to identify the most prominent emission lines in the different regions, that can then be used for imaging and further inference on the localized physical state of the plasma \citep[see][]{Paggi12,Fabbiano18b,Maksym19}. 
We then fitted the spectra with a range of physical models to investigate the most probable mechanisms contributing to the emission in each region \citep[e.g.][]{Bianchi06,Fabbiano18a,Jones20}. The models fitted were:
\begin{enumerate}

\item A leaky absorber model for the nuclear spectrum, to a facilitate comparison with previous work \citep{Levenson06,Fabbiano18a};
\item A simple reflection model (PEXRAV), for the reflection of X-rays by the cold material of the accretion disk and the torus \citep{Magdziarz95}, and a more complex reflection model \citep[PEXMON;][]{Nandra07}, that self-consistently generates iron and nickel emission lines (see Appendix~\ref{sec:procedure1} for details).
\item Photoionization models \citep[CLOUDY;][]{Ferland98}, consisting of a grid of values produced with the CLOUDY c08.01 package. The variables in CLOUDY are the ionization parameter\footnote{$\rm U \simeq \int _{\nu _R} ^{+ \infty} L_{\nu} d \nu / 4 \pi r^2 c n_e$ with r the distance of the gas from the source, $\rm L_{\nu}$ the ionizing luminosity, $\rm \nu _R$ the Rydberg frequency, and $\rm n_e$ the electron density.} (log~U=[-3.00:2.00] in steps of 0.25) and hydrogen column density (log $\rm N_H$=[19.5:23.5] in steps of 0.1) through the irradiated slab of gas, where the assumption is that the irradiation is from photons produced in the AGN;
\item Optically thin thermal emission \citep[APEC;][]{Foster12} that can result from the thermalization of the ISM after being collisionally ionized by interaction with the radio jet or winds from either the nucleus or from star forming regions;
\item Thermal emission originating directly from the shock fronts \citep[PSHOCK;][]{Borkowski01}. This model assumes a plane-parallel shocked plasma with constant post-shock electron and ions temperature, element abundances, and ionization timescale, providing a useful approximation for supernova remnants, but more generally for all cases in which X-ray emission is produced in a shock front.

\end{enumerate}

Following the procedure in previous work \citep[e.g.][]{Fabbiano18a}, the data were fitted first with a single model, and then with combinations of an increasing number of models. To establish the goodness of fit, we performed both standard statistical tests and also examined the residuals in the different spectral ranges. Details of the procedures and results are given in Appendix~\ref{sec:procedure1} for the nuclear region, and Appendix~\ref{sec:procedure2} for the extended cone and cross-cone regions.

The results of these spectral fits show that the physical state of the emitting ISM is complex (see Appendix~\ref{sec:procedure1},~\ref{sec:procedure2}). Typically, multiple-component models are needed to account for the various spectral features, including both photoionized and thermal components. In particular a range of ionization parameters is suggested by the data [$log~U \sim -2.8,~1.9$], and also a range of temperature [$kT \sim 0.3,~2.9 ~keV$]. Given that we are extracting spectra from large physical regions, this complexity is not surprising. The ISM is expected to have a range of cloud densities, and temperature \citep[e.g., see the case of ESO428-G014,][]{Fabbiano18b}.

The nuclear spectrum shows a strong feature from a blend of Mg XII and Si XIII lines that require photoionization. However, a single CLOUDY component, in addition to the PEXMON AGN model does not fit well the entire spectrum. Two-component models are required, either two CLOUDY components, or a CLOUDY component plus a thermal (APEC or PSHOCK component).
For the extended emission, a hard reflection power-law and Fe Ka emission (PEXMON) is needed to fit the NW and SE cone spectra, in addition to both photoionization and collisional ionization. The cross-cone spectrum does not have an intrinsic hard emission: the hard X-ray continuum and Fe Ka line can all be explained in terms of ``spillover'' of the nuclear spectrum, due to the PSF wings (Appendix~\ref{sec:procedure2}).

In Section~\ref{sec:discussion} below we discuss the possible scenarios allowed by the range of spectral fit results, and we refer back to these results as needed.

\section{Discussion} \label{sec:discussion}

Deep Chandra observations of nearby obscured and Compton Thick (CT; $log(N_H/cm^2) > 24.5$) AGNs are significantly improving our understanding of the nucleus and its immediate surrounding, and of the AGN-host interaction in gas-rich galaxy disks (e.g., NGC4151, \citealt{Wang11a,Wang11b,Wang11c}; Mkn 573, \citealt{Paggi12}; NGC 3393, \citealt{Maksym19}; ESO 428-G014, \citealt{Fabbiano18a,Fabbiano18b,Fabbiano19}; NGC 7212, \citealt{Jones20}). The deep Chandra observations of IC 5063 presented in this paper give us another detailed case study of these phenomena.   
A previous Chandra study reported extended soft X-ray emission from a kpc-size [O III] ionization bi-cone in the host galaxy disk of IC 5063 \citep{Guijarro17}. This discovery motivated our deep Chandra ACIS observations, which have revealed extended emission both in the bi-cone direction, and in the perpendicular (``cross-cone'') direction (Section~\ref{sect:imradprof}). We have characterized the spectral properties of this emission and also of that of the bright nuclear point-like source (Sect.~\ref{Spec}; and see Appendix~\ref{sec:procedure1}, \ref{sec:procedure2} for details). Below we discuss our results and their implications.

We first discuss the nuclear emission and compare our results with previous X-ray studies of this AGN (Section~\ref{sec:DiscNuc}). We then discuss the X-ray emission of the kpc-size bi-cone (Section~\ref{sec:DiscBiCone}), which is the region of direct interaction of the AGN with the host disk; the soft emission in the cross-cone region, which extends above and below the host disk (Section~\ref{sec:DiscCC}); and the energy-dependence of the diffuse emission (Section~\ref{sec:DiscED}), both in the cone and in the cross-cone. Finally, we discuss some evidence of possible interaction of the radio jets with the hot ISM (Section~\ref{sec:DiscJet}).

\subsection{The Nuclear Emission} \label{sec:DiscNuc}

The nuclear spectrum exhibits both strong hard ($>$3~keV) X-ray emission and a soft excess at lower energies. These characteristics are suggestive of direct nuclear coronal emission in a leaky absorber model \citep{Reichert85} with a high covering fraction of $99.2 \pm 0.2 \%$, plus a reflection component. Alternatively, the spectrum can also be fitted (albeit formally less well, Table~\ref{table:parNuc}) by the model used in previous work on IC 5063 by \cite{Vignali97} and \cite{Tazaki11}. This model fits the hard part of the X-ray spectrum with a power law and reflection PEXRAV component associated with an absorption model, and fits the soft excess with a simple power law. Although both models require a reflection and two power law components, in the leaky absorber model the two power laws represent a single partially absorbed continuum, while in the approach used by Vignali et al. and Tazaki et al. the two components are decoupled, with quite different photon indices and intrinsic absorptions.

At energies $>$3 keV we detect a 6.4 keV neutral Fe K$\alpha$ line (14$\sigma$), with an $EW \sim 178_{-41}^{+55}~eV$ consistent both with \cite{Vignali97} and \cite{Tazaki11} for IC 5063. This low EW is typical for Seyfert 1s ($<$0.5 keV), while Seyfert 2s usually exhibit Fe K$\alpha$ $EW \sim0.5-2 ~keV$ \citep{Singh11,Shu11}. However, some Seyfert 2s have similarly low EW to IC 5063, e.g. Mrk 348 \citep[$EW\sim 40~eV$,][]{Singh11}, Mrk 477 \citep[$EW \sim 100-300~eV$,][]{HernandezGarcia15}. These low EW may be indicative of an ``unobscured'' Seyfert 2 galaxy \citep[see e.g.][]{Brightman08,Bianchi12}, in which although lacking the broad-line region, the hard X-ray emission appears unabsorbed. In our case \cite{Inglis93} detected broad emission lines in polarized flux, thus suggesting a more complex structure of the nucleus in which the broad-line region is obscured, while the primary X-ray emission can escape unattenuated. 
If instead of calculating the EW from the total continuum, we measure it with respect to the PEXMON reflected continuum component (see Table \ref{table:parNuc}), we obtain EW$\sim 1.96_{-0.62}^{+1.23}~keV$, that is consistent with the standard scenario of an emission line reflected from circumnuclear material \citep{Smith93, Singh11}.

The most prominent feature below 3~keV is at $\sim$1.8~keV. We associate this feature with a blend of the Mg XII (1.745~keV; $\sim$5$\sigma$) and Si~XIII (1.865~keV; $\sim$2$\sigma$) transitions, which are an indication of photoionized emission \citep{Koss15}.
\cite{Liu19} suggested that the Si~XIII line can be also related to outflowing hot gas.

\begin{figure}[t]
   \begin{center}
   \includegraphics[height=0.3\textheight,angle=0]{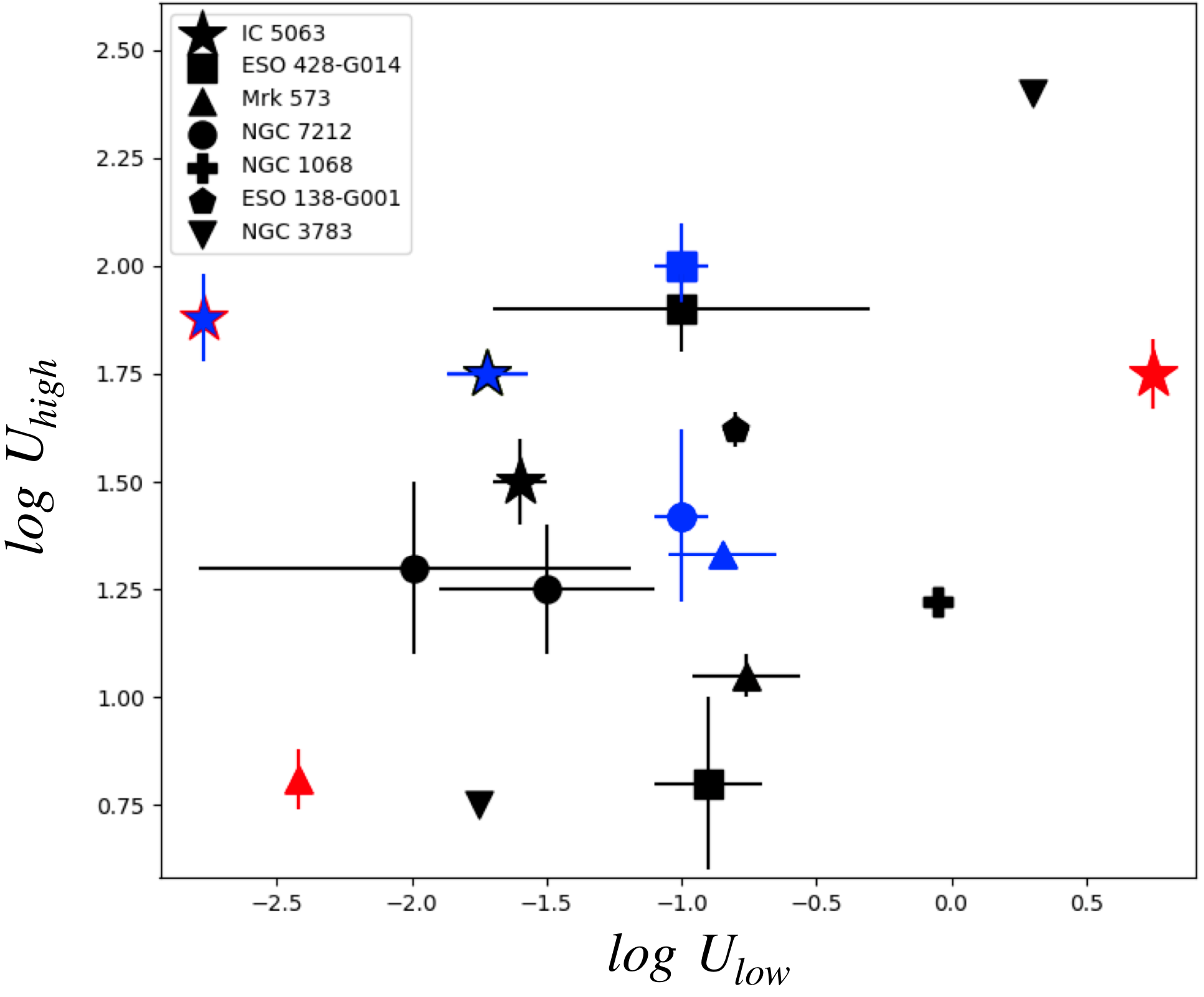}
   \caption{Lower and higher photoionization parameters of the best-fit models obtained for Seyfert galaxies in the literature, compared with our results for IC 5063 (large stars). Nuclear (black); bi-cone (blue); cross-cone (red). The blue stars show a red and a black edge indicating the values for the NW and SE cone, respectively. The blue marks by ESO428-G014 \citep{Fabbiano18a} and NGC 7212 \citep{Jones20} represent values obtained from the spectral fits in the extended annular region, where, however, the emission along the cones dominates. We included the following Seyferts: Mrk 573 \citep{Paggi12}, NGC 1068 \citep{Kraemer15}, ESO-138-G001 \citep{DeCicco15} and NGC 3783 \citep{Kaspi02,Blustin02}.}
   \label{image:U1U2}
   \end{center}
\end{figure}

Multi-component fits to photoionization and thermal emission (Table~\ref{table:parNuc}) show that the nucleus is dominated by a low-photoionization ($log U \sim -1.6$), high column density ($log[N_H/cm^2] \sim 23.5$) gas component, that allow us to reproduce prominent emission lines at $\sim 1.7-1.9$ and $\sim 2.3~keV$. A second, high photoionization component ($log U \sim 1.5$), with similarly high column density absorption can model the soft excess continuum. 
In Fig.~\ref{image:U1U2} we report the largest and lowest photoionization parameters used in the same model to fit the spectra of different regions of other Seyferts. The ionization parameters we estimate for the nuclear region of IC 5063 are compatible with those reported in literature. Alternatively, a collisional component with temperature $kT \sim 1.30~keV$ or a shock model with $kT\sim 2.87~keV$ are equally acceptable (see Table~\ref{table:parNuc}).

The nuclear region used for the extraction of the spectrum (Section \ref{Spec}) encloses the unresolved nucleus (which dominates the hard emission), the inner regions of the [OIII]/H$\alpha$ ionization bi-cone \citep{Colina91}, and most of the region of interaction with the radio jets \citep{Morganti98}. This could explain the complexity of the spectral parameters. In particular, the two temperatures may be consistent with shock velocities of $\sim 800~km~s^{-1}$ and $\sim 1200~km~s^{-1}$, respectively (see Table~\ref{table:phypar}), assuming $v_{shock}= \sqrt{16 kT/3 \mu}$ \citep{Wang14}. Both the collisionally and shock ionized models predict ISM densities $n_e \sim 0.1~cm^{-3}$ (assuming a filling factor $\eta =1$), which are an order of magnitude less than the ISM density estimated in the nucleus of ESO428-G014 \citep[$n_e \sim4~cm^{-3}$,][Table~9]{Fabbiano18a}, and similar to those in the outer optical arcs in Mrk 573 \citep[$n_e \sim 0.14~cm^{-3}$,][]{Paggi12}. This result is puzzling, as these low densities are inconsistent with those expected in gas-rich disk galaxies. They may be explained as an overestimate of the filling factor due to clumpiness in the ionized gas, or to our seeing the emission coming from a thin skin of hot gas above the galaxy plane, possibly due to a AGN wind host - disk interaction \citep{Maksym19}. A third possibility is that the low densities could be real and due to strong AGN winds evacuating the ISM from the galaxy in these locations.
Table~\ref{table:phypar} lists how the physical parameters depend on the filling factor $\eta$. In the table we give values for $\eta=$1. Of note are the cooling times of $\sim 10^7 yrs$, which could indicate a transient phenomenon as also indicated in \cite{Mukherjee18} simulations, which found a cooling time lower than the dynamical time in the core ($log~t_{cool}/t_{dyn} \leq -2$).

\begin{table*}[t]
\begin{center}
\caption{Physical parameters of the collisionally and shock ionized gas in all the regions.}
\label{table:phypar}
\begin{tabular}{c|c|c|c|c|c|c|c|c|c}
\toprule
Region & Model [tot]$^a$ & $V \eta$ & $L_{0.3-10~keV}$ & $n_{e} \eta^{-1/2}$ & $E_{th} \eta^{1/2}$ & $P_{th} \eta^{-1/2}$ & $t_{cool} \eta^{1/2}$ & $M \eta^{1/2}$ & $v_{shock}$ \\
  &  & [$10^{65} cm^3$] & [$10^{39}~\frac{erg}{s^1}$] & [$cm^{-3}$] & [$10^{55}~ erg$]  & [$10^{-12} ~\frac{dyne}{cm^{2}}$] & [$10^7~yr$] & [$10^6~M_{\odot}$] & [$km~s^{-1}$]  \\[2pt]
\hline
Nucleus & APEC [AC]   & 0.14 & 3.2 & $0.11_{-0.06}^{+0.07}$    & $0.5_{-0.2}^{+0.4}$ & $248 \pm 114$     & $0.5_{-0.2}^{+0.4}$ & $1.2_{-0.7}^{+0.8}$    & $779 _{-301}^{+ 255}$ \\
        & PSHOCK [PC] & --   & 6.2 & $0.12_{-0.05}^{+0.06}$    & $1.4_{-0.3}^{+1.7}$ & $671 \pm 162$     & $0.7_{-0.2}^{+0.9}$ & $1.4_{-0.5}^{+0.7}$    & $1200_{-940}^{+1029}$ \\
NW Cone & APEC [AC]   & 4.10 & 1.4 & $0.012_{-0.006}^{+0.006}$ & $1.4_{-0.5}^{+1.1}$ & $22.9 \pm 8.4$    & $3.1_{-1.1}^{+2.4}$ & $4.0_{-2.0}^{+2.0}$    & $712 _{-255}^{+ 368}$ \\
SE Cone & APEC [AC]   & 4.10 & 1.0 & $0.011_{-0.007}^{+0.009}$ & $1.9_{-0.9}^{+1.9}$ & $30.4 \pm 14.9$   & $6.1_{-3.0}^{+6.2}$ & $4.0_{-2.5}^{+3.1}$    & $826 _{-309}^{+ 401}$ \\
Cross-Cone & PSHOCK [PP] & 27.9 & 3.9 & $0.018_{-0.010}^{+0.008}$ & $9.7_{-3.0}^{+6.3}$ & $23.3 \pm 7.1$ & $7.8_{-2.4}^{+5.1}$ & $41 _{-24}^{+19}$ & $ 584_{-309}^{+245}$ \\
           & PSHOCK [PP] &  --  & 4.2 & $0.006_{-0.002}^{+0.002}$ & $5.1_{-1.5}^{+2.4}$ & $12.2 \pm 3.6$ & $3.8_{-1.1}^{+1.8}$ & $14 _{- 6}^{+4}$  & $ 719_{-235}^{+283}$ \\
           & PSHOCK [PC] &  --  & 3.7 & $0.005_{-0.003}^{+0.003}$ & $7.6_{-2.7}^{+6.4}$ & $18.1 \pm 6.3$ & $6.5_{-2.2}^{+5.5}$ & $13_{- 7}^{+ 7}$  & $ 926_{-413}^{+516}$ \\
           & PSHOCK [PA] &  --  & 5.7 & $0.006_{-0.002}^{+0.002}$ & $10 _{-2  }^{+8.1}$ & $24.7 \pm 4.7$ & $5.7_{-1.0}^{+4.5}$ & $14_{- 5}^{+ 5}$  & $1044_{-465}^{+665}$ \\
           & APEC   [PA] &  --  & 1.7 & $0.005_{-0.003}^{+0.004}$ & $2.0_{-0.5}^{+2.2}$ & $ 4.8 \pm 1.3$ & $3.7_{-1.0}^{+4.0}$ & $11_{- 7}^{+ 9}$  & $ 521_{-317}^{+274}$ \\
           & APEC   [AC] &  --  & 2.2 & $0.006_{-0.003}^{+0.003}$ & $6.2_{-2.6}^{+3.8}$ & $14.8 \pm 6.2$ & $8.8_{-3.7}^{+5.4}$ & $14_{- 8}^{+ 7}$  & $ 786_{-274}^{+255}$ \\
           & APEC   [AA] &  --  & 3.6 & $0.008_{-0.003}^{+0.003}$ & $7.7_{-2.3}^{+3.4}$ & $18.5 \pm 5.6$ & $6.7_{-2.0}^{+3.0}$ & $18_{- 7}^{+ 7}$  & $ 779_{-213}^{+200}$ \\
           & APEC   [AA] &  --  & 3.6 & $0.008_{-0.005}^{+0.005}$ & $2.0_{-1.0}^{+1.8}$ & $ 4.7 \pm 2.3$ & $1.7_{-0.8}^{+1.6}$ & $18_{-12}^{+12}$  & $ 388_{-158}^{+200}$ \\[5pt]
\toprule
\end{tabular}
{\raggedright \textbf{Notes.}  $^a$ "Model" represents the template from which we derived the values and the complete configuration of the model to which it belongs is "tot", where A=APEC, C=CLOUDY, P=PSHOCK. \par}
{\raggedright We assumed a filling factor $\eta=1$. We, therefore, show all the parameters as function of the filling factor. \par}
{\raggedright Errors are quoted at $1 \sigma$ significance. \par}
\end{center}
\end{table*}

\begin{table}[t]
\begin{center}\footnotesize{
\caption{Percentage counts relative to the Nucleus in image (left) and PSF (right) and excess counts over the $Chandra$ PSF (bottom), in the off-nuclear regions at different energy bands.}
\label{table:excess}
\begin{tabular}{c|c c|c c|c c|c c|c c|c c|c c}
            \toprule
\hline
\multicolumn{1}{c|}{} & \multicolumn{14}{c}{Energy bands [keV]} \\
\hline
Region & \multicolumn{2}{c|}{0.3-1.5} & \multicolumn{2}{c|}{1.5-3.0} & \multicolumn{2}{c|}{3.0-4.0} & \multicolumn{2}{c|}{4.0-5.0} & \multicolumn{2}{c|}{5.0-6.0} & \multicolumn{2}{c|}{6.1-6.6} & \multicolumn{2}{c}{7.0-8.0} \\
\hline
\hline
NW cone  &          32.7 & 0.4 & 14.6  & 0.7 &  2.5 & 1.1 & 1.8 & 1.2 & 1.4 & 1.1 & 2.3 & 1.0 & 1.2 & 0.5  \\ 
& \multicolumn{2}{c|}{442}  &  \multicolumn{2}{c|}{317}  &  \multicolumn{2}{c|}{59}  &  \multicolumn{2}{c|}{46}  &  \multicolumn{2}{c|}{18}  &  \multicolumn{2}{c|}{44}  &  \multicolumn{2}{c}{11}   \\
SE cone  &         22.8 & 0.4 & 13.7  & 0.7 &  3.2 & 1.1 & 2.0 & 1.2 & 1.5 & 1.1 & 1.9 & 1.0 & 1.3 & 0.5  \\ 
& \multicolumn{2}{c|}{307}  &  \multicolumn{2}{c|}{295}  &  \multicolumn{2}{c|}{89}  &  \multicolumn{2}{c|}{65}  &  \multicolumn{2}{c|}{33}  &  \multicolumn{2}{c|}{29}  &  \multicolumn{2}{c}{14}  \\
NE cross-cone &   21.0 & 0.7 & 5.5   & 1.3 &  2.7 & 2.0 & 2.2 & 2.3 & 2.3 & 2.1 & 2.3 & 2.0 & 2.0 & 0.9  \\
&   \multicolumn{2}{c|}{278}  &  \multicolumn{2}{c|}{93}   &  \multicolumn{2}{c|}{26}  &  \multicolumn{2}{c|}{--}  &  \multicolumn{2}{c|}{15}  &  \multicolumn{2}{c|}{12}  &  \multicolumn{2}{c}{17} \\  
SW cross-cone &   19.7 & 0.7 & 4.5   & 1.4 &  2.0 & 2.1 & 2.1 & 2.3 & 2.2 & 2.1 & 2.5 & 1.8 & 2.4 & 0.8 \\ 
&   \multicolumn{2}{c|}{259}  &  \multicolumn{2}{c|}{70}   &  \multicolumn{2}{c|}{--}  &  \multicolumn{2}{c|}{--}  &  \multicolumn{2}{c|}{7}   &  \multicolumn{2}{c|}{21}  &  \multicolumn{2}{c}{24} \\
             \toprule
\end{tabular}
{\raggedright \textbf{Notes.}  "--" indicates absence of excess counts. \par}
}
\end{center}
\end{table}

\subsection{The Ionized Bi-cone} \label{sec:DiscBiCone}

Figs.~\ref{image:regions} and \ref{image:rp1} clearly show the presence of X-ray emission stretching E-W along the bi-cone direction out to $\sim$2~kpc from the nucleus. This emission is particularly prominent in the soft band ($<$3~keV). Extended soft ($<$3~keV) X-ray emission has been observed in Seyfert 1.5 - 2 galaxies, spatially correlated with the [OIII] emission \citep{Bianchi06,Bianchi10,Levenson06}, suggesting a common origin largely due to photoionization, and partially to collisional ionization, of circumnuclear clouds \citep{Wang11c,Matt13}. Prominent emission lines are found in the spectrum below 3.0 keV, typically observed in CT Seyfert \citep{Wang11a,Fabbiano18a,Maksym19,Jones20}. These lines include Ne X, Mg XI, Si XIII, Si XIII, S XV and the Fe L complex. Given the ACIS spectral resolutions these lines are typically blended in the observed spectra (Table~\ref{table:parExtReg} in Appendix~\ref{sec:procedure2}). 
The best-fit models of the bi-cone spectra need to include, at least, one photoionization component.
Focusing on the best-fit models consisting of 2 photoionized phases, Fig.~\ref{image:U1U2} shows that the NW cone photoionization parameters put it at the periphery of the Seyfert 2 distribution, towards lower $U_{low}$. In this case, the presence of two quite different photoionization phases in the bi-cone direction could be justified by the X-shape morphology of the ionization cones \citep{Colina91}, implying the co-existence of two net separated regions differently illuminated by the central AGN.
In both cone regions the higher photoionization component can be replaced by a collisionally ionized component with temperature $kT \approx 1.2~keV$.

The resulting average ISM density is $\sim 0.01~cm^{-3}$ for both sides, where we are considering average density of thermal gas in a spherical shell with angle 60 degree and from $\sim 0.5$ to $\sim 3.4~kpc$.
As in the nucleus, these densities are lower than those reported for other Seyfert 2s \citep{Paggi12,Fabbiano18a,Maksym19}. This suggests more clumpiness of the hot gas in IC 5063. Since these densities are $\sim$1/10 of those observed in the nuclear region, the cooling times are correspondingly longer $\sim 3.0 \times 10^8 ~yr$ (Table~\ref{table:phypar}).

Figs.~\ref{image:rp1} and \ref{image:rp2} show that the bi-cone emission is present also at energies $>$3 keV. Harder extended components (both in the continuum emission above $\sim$3~keV and the neutral 6.4 keV Fe K$\alpha$ line) have been detected with Chandra in AGNs \citep[see][]{Bauer15,Fabbiano17,Jones20,Jones21,Ma20}. The bi-cone spectra clearly show a roughly flat hard X-ray continuum in IC 5063, which needs to be fitted with a reflection component, whose normalization is weaker (by a factor 2) and the photon index is steeper ($\Gamma = 2.7$) compared to that of the nuclear spectrum ($\Gamma= 1.45$). 
In agreement with \cite{Fabbiano17}, we suggest that the steeper hard X-ray continuum seen in extended regions  implies this is due to the scattered and/or fluorescent intrinsic emission escaping unattenuated in the bi-cone direction.

Table~\ref{table:excess} gives both the percentages of the counts with respect to the nuclear region, and the excess counts over the PSF, in the merged image and in the different sectors (as defined in Section~\ref{Spec}) and energy bands. At energies $<$3 keV the percentage of counts in the bi-cone regions is significantly larger than the percentage predicted for the nuclear spillover (Appendix~\ref{sec:NucSpillover}). Of these excess counts 66$\%$ is from within the bi-cone sectors, which represent 1/3 of the total external area.  

\begin{figure}[t]
   \begin{center}
   \includegraphics[height=0.26\textheight,angle=0]{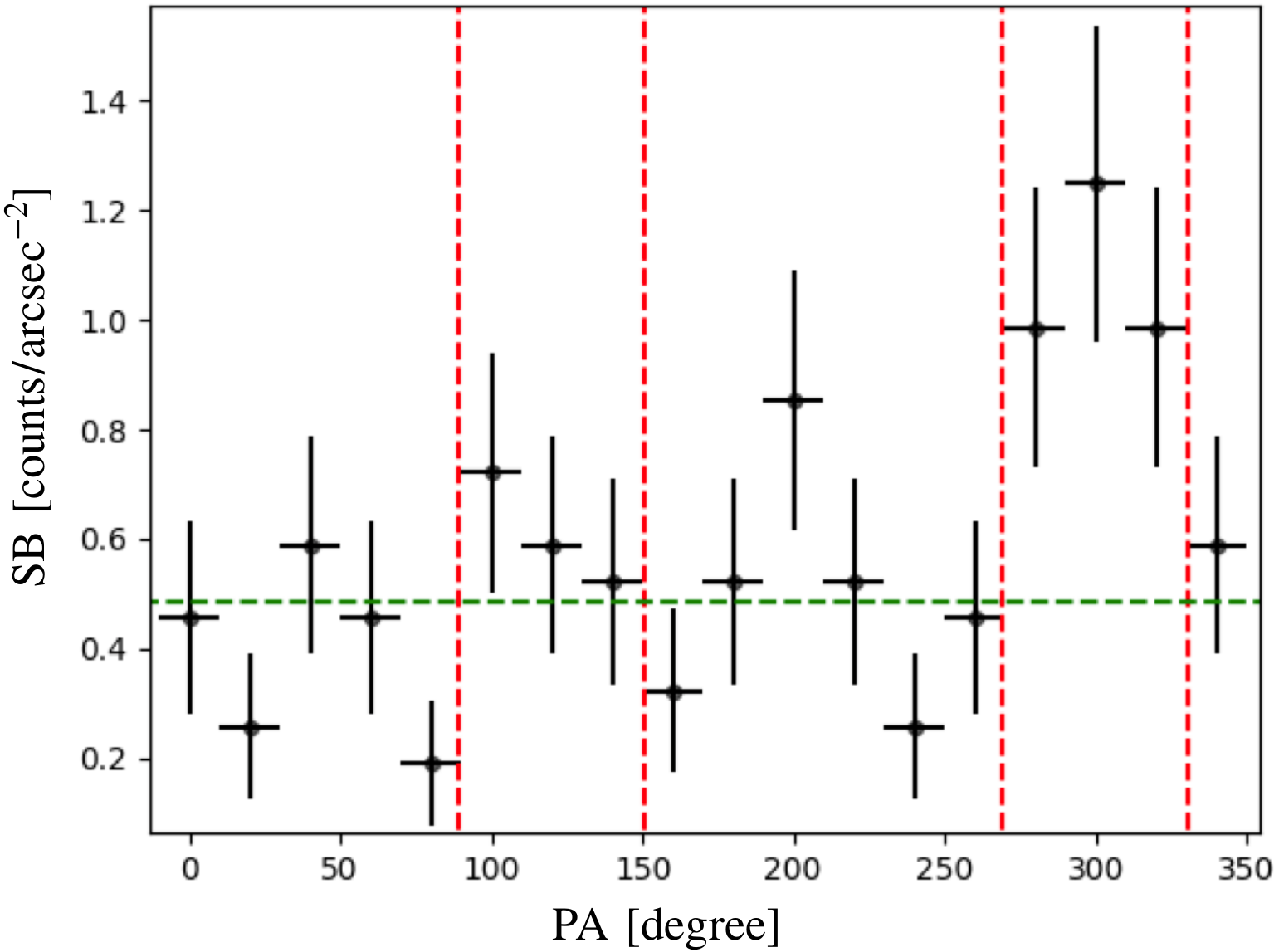}
   \caption{Surface brightness (in $\rm counts/arcesc^2$) azimuthal profile from 0.5 to 1.2 kpc in the 6.1-6.6 keV image (Fe K$\alpha$). Red dashed vertical lines separate the bi-cone and cross-cone sectors, and the green dashed line indicates the average of the surface brightness, which is estimated excluding the points in the NW sector.}
   \label{image:PAFeKa}
   \end{center}
\end{figure}

A neutral iron emission line is observed in both the bi-cone spectra, and most of this feature is modelled by a reflection PEXMON component \citep[e.g.][]{George91,Matt91} with solar abundance, while a small ($\sim$16 $\%$) contribution is reproduced with photoionization CLOUDY model.
However, in the energy band 6.1-6.6 keV, which is dominated by this Fe K$\alpha$ emission, 60$\%$ (44/73) of the counts in the 2$''$-15$''$ annulus is in the NW cone. This excess of Fe K$\alpha$ emission corresponds to the protrusion at a projected distance of 5$''$ (1.2~kpc) in Fig.~\ref{image:rp2} at these energies. This protrusion of the neutral iron emission in the NW cone is well visible in the azimuthal surface brightness profile in Fig.~\ref{image:PAFeKa}, showing a 2-3$\sigma$ significance. 
The Fe K$\alpha$ emission spatially correlated with the most intense radio hot-spot suggests the presence of high density clouds in this region. 
These clouds may fluoresce because of the interaction with AGN photons escaping in the jet direction. Alternatively, X-rays may also be produced locally by jet induced shocks and interact with the clouds (see Sect.~\ref{sec:DiscJet}).

\subsection{The Cross-Cone emission} \label{sec:DiscCC}

\begin{figure}[t]
   \begin{center}
   \includegraphics[height=0.34\textheight,angle=0]{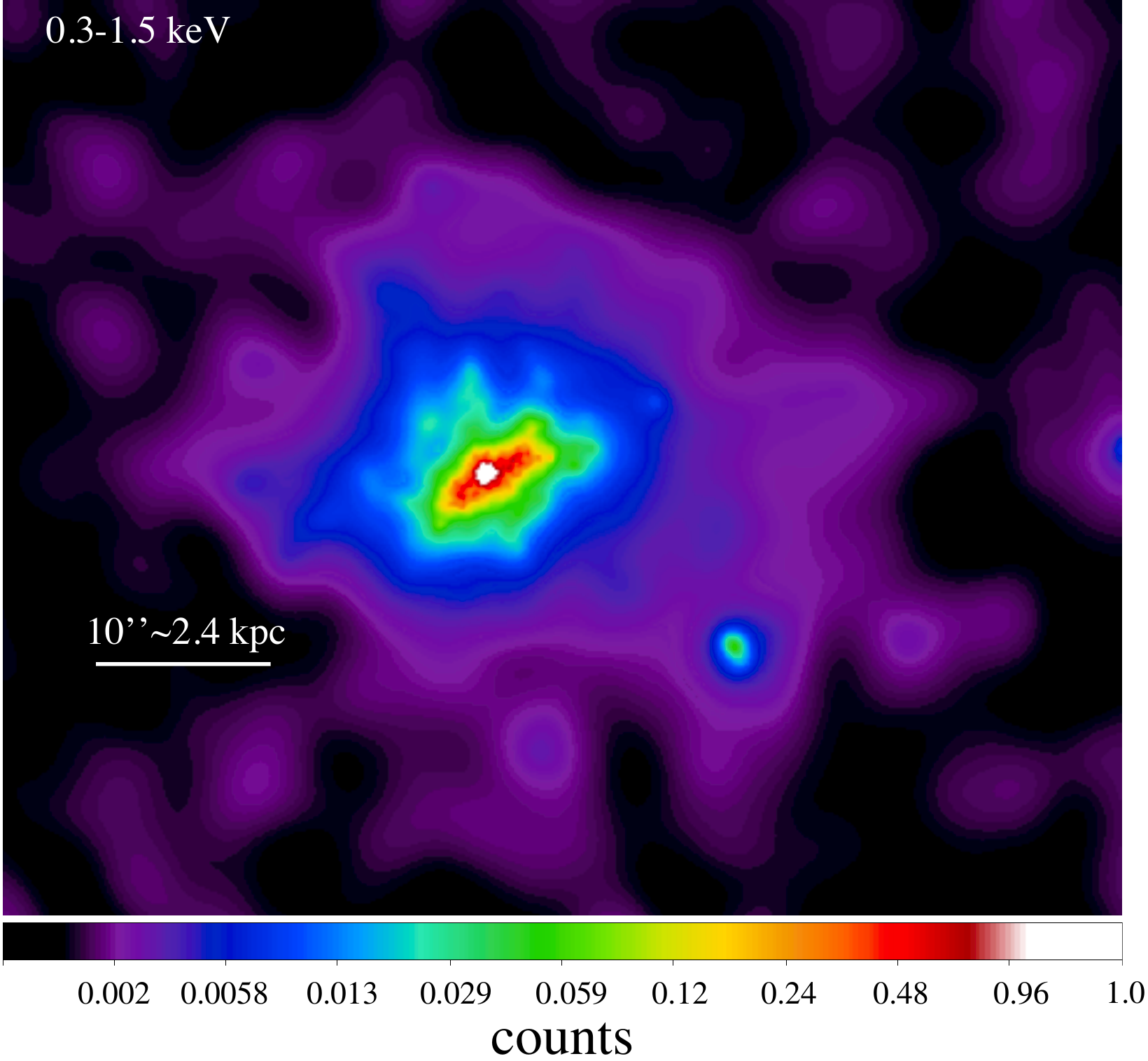}
   \caption{0.3-1.5 keV image, adaptively smoothed with Gaussian kernels ranging from 0.5 to 30 image pixel (1/8 of instrumental pixel) in 30 iterations and 10 counts under the kernel. The color bar is in counts per image 1/8 subpixel.}
   \label{image:TorusSoft}
   \end{center}
\end{figure}

Fig.~\ref{image:rp1} and Table~\ref{table:excess} show significant extended emission in the cross-cone region. This emission is more prominent at energies $<$1.5~keV and extends out to a radius of $\sim 3~kpc$ from the nucleus. Fig.~\ref{image:TorusSoft} shows the 0.3-1.5 keV image, adaptively smoothed with Gaussian kernels ranging from 0.5 to 30 image 1/8 of instrumental pixel in 30 iterations and 10 counts under the kernel. The color scale has been chosen to minimize the visual impact of the nuclear and bi-cone emission.

Extended soft X-ray emission, perpendicular to the bi-cone axis, has been observed in other AGNs \citep{Wang11a,Paggi12,Fabbiano18a,Maksym19,Jones20,Jones21}. This emission is not expected from the classical AGN unification paradigm \citep{Antonucci93}. A possible explanation could be that the nuclear torus may be porous and allows part of the nuclear photoionizing continuum to escape \citep{Nenkova08}. The volume of the cross-cone is 4 times the volume of the bi-cone, and photons with energies 0.3-1.5~keV in the cross-cone are 73$\%$ of those in the bi-cone. Therefore, following \cite{Fabbiano18a}, the transmission of the torus in the cross-cone direction is 20$\%$, twice that found for ESO428-G014. Alternatively, this emission may be due to hot outflows from the galactic disk of IC 5063, caused by the interaction of the jet with the ISM, as predicted by the simulations of \cite{Mukherjee18}. This scenario would be in agreement with the conclusions in \cite{Maksym20a}, \cite{Maksym20} and \cite{Venturi21}, that found suggestions for lateral outflows as a consequence of the radio jet-ISM interaction, detecting ``dark rays'' in HST near-infrared data perpendicular to the galaxy disk as suggestion for the presence of large-scale dust, a low-ionization [SII] emission loop and high velocity dispersion of H$\alpha$ and [OIII] emission lines in the cross-cone area, respectively.
The best phenomenological model of the cross-cone spectrum (see Section~\ref{sec:PhModCC}) consists of a steep $\Gamma = 3$ power law plus a flat reflection component ($\Gamma < 1.2$).
The spectrum shows an excess at 0.9-1 keV, which could be due to blended Fe-L emission lines (e.g. Fe XXI [1.009 keV]) and may also include Ne IX [0.915 keV] and/or Ne X [1.022 keV] lines. It does not exhibit the strong emission lines at $\sim 1.8~keV$ or $\sim 2.3~keV$ observed in the nuclear and bi-cone spectra, which can only be reproduced by photoionization models. In the cross-cone region, we find no evidence for a neutral iron line at 6.4 keV. 

The cross-cone spectrum can be fitted with a mix of any 2 components among photoionization, collisional and shock ionization models. 
The principal photoionization component has $log U  \sim 0.8$ and very low column densities $log(N_H/cm^{2}) < 19.7$.
For the fit with two components of photoionized gas, the second phase has a higher ionization parameter $log U \sim 1.7$ and $log(N_H/cm^{2}) < 20.6$.  Fig.~\ref{image:U1U2} shows that both the ionization parameters for the cross-cone spectrum are larger than those found by \cite{Paggi12} for the cross-cone of Mrk 573. These high values may suggest that the emission is not due to photoionization from AGN photons escaping from a leaky torus but instead is prevalently thermal.

The temperatures we find for shock ionized gas range from 0.5 to 3$~keV$, while the collisionally ionized gas temperatures are between 0.3 and 1.4$~keV$. The normalization values of the thermal models imply nominal densities of the emitting gas $n_e \sim 0.006~cm^{-3}$ (see Table~\ref{table:phypar}), a factor $\sim$2 lower than those in the bi-cone region. Given the large volume, the range of masses ($\sim 1-4 \times 10^7 M_{\odot}$) in the cross-cone is larger than what we observed in the bi-cone regions. Collisionally and/or shock ionized gas would be consistent with the predictions of the \cite{Mukherjee18} simulations, in which in the later stage of the jet-ISM interaction, gas at $T \sim 10^7~K$ would be swept away from the disk in the form of large-scale filamentary winds with velocities $>500~km~s^{-1}$. The observed gas densities, however, are a factor of 10 lower than expected for perpendicular filaments in \cite{Mukherjee18} simulations. Alternatively, if we assume that this thermal gas has a density $n_e \sim 0.1-0.3~cm^{-3}$, as predicted by \cite{Mukherjee18} (see their Figs.~2 and 4), we derive a filling factor $\eta < 0.01$. This is exactly consistent with the filling factor lower than 1$\%$ estimated by \cite{Oosterloo17} in ALMA observations of molecular gas (CO transitions). Hence, we suggest that the filling factor is far from unity and that the hot and cold phase of the gas, in the inner jet-affected regions, exhibit a similar clumpiness. 
The cohabitation of the hot X-ray emitting and molecular CO emitting gas in the jet-disturbed circumnuclear regions is observed in other objects \citep[e.g.][]{Feruglio20,Grossova19}. That these two different gas phases share the same clumpiness suggests that the radio jet impact has a similar fragmentation effect on these two phases. According to a typical scenario, the high pressure due to the jet compresses and accelerates fast, shock-driven, lateral outflows, increasing the density and temperature of the molecular gas \citep{Oosterloo17} and producing both soft X-ray photons by shock and hard X-ray photons by scattering by dense clouds \citep{Fabbiano17}. 
However, the question remains whether this cold phase was already present in the circum-nuclear medium, mixed with the hot phase gas, before the impact with the radio jet, or whether it represents gas that, for some reason, has cooled down or has not warmed up. A careful spatial analysis comparing cold molecular gas and its hot phase in the jet-affected region is planned to get a more complete picture.
In conclusion, emission from shocked and/or collisionally ionized gas is preferred over photoionized gas for the cross-cone region in IC 5063.

\begin{figure}[t]
   \begin{center}
   \includegraphics[height=0.22\textheight,angle=0]{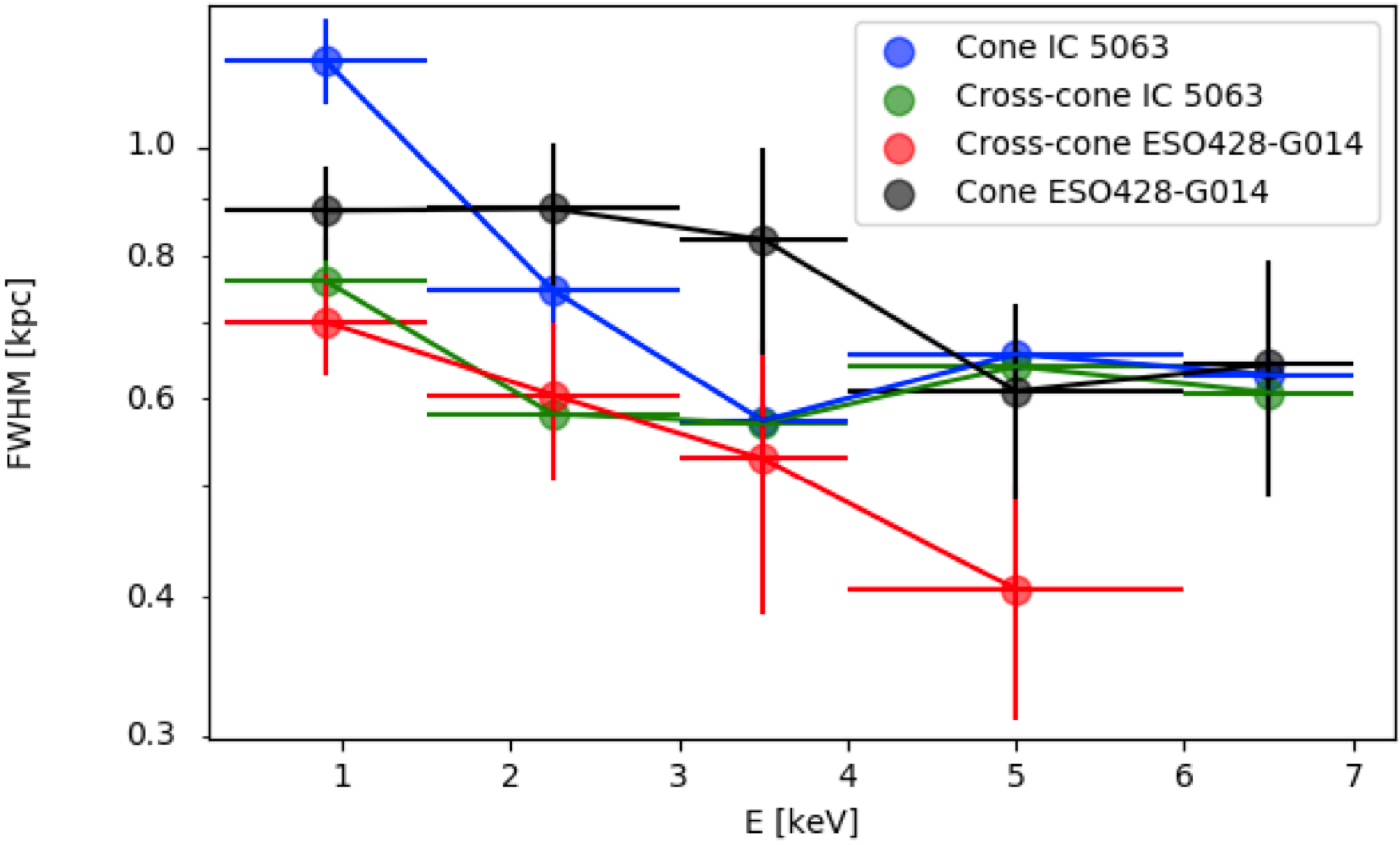}
   \caption{FWHM (units of kpc) of the radial profiles in Figs.~\ref{image:rp1} and \ref{image:rp2} (blue, green) and for ESO428-G014 in \cite{Fabbiano18a} (black, red) in a log space, as a function of the energy band, along the bi-cone and cross-cone direction.}
   \label{image:ExtEneESO}
   \end{center}
\end{figure}

\subsection{Energy-Dependence of the Extent} \label{sec:DiscED}

\cite{Fabbiano18a} noticed that the extent of the large-scale kiloparsec-size emission of the obscured AGN ES0 428-G014 decreases with increasing photon energy, and suggested that this effect may be related to a more central concentration of the denser molecular clouds responsible for the reflection and scattering of the higher energy nuclear photons in the galaxy disk. More recently, \cite{Jones21} have confirmed this effect for a sample of five CT AGNs studied with Chandra. We find a similar behavior in IC 5063, as shown both by Table~\ref{table:excess} and Fig.~\ref{image:ExtEneESO}, which compares the large-scale extent of the emission of IC 5063 in different energy bands with that of ESO428-G014, following \cite{Fabbiano18a}.
Fig.~\ref{image:ExtEneESO} shows that at energies $<$4 keV in both AGNs the extent is larger in the cone direction. The decrease in the extent of the X-ray emission, at energies $<$4 keV, along the cross-cone direction in IC 5063 is similar to that found for ESO428-G014. The extent of the bi-cone regions in IC 5063 is a steeper function of energy than in ESO428-G014. 

\cite{Jones21} associate the energy dependence of the extent of the bi-cone with its inclination relative to the galactic disk. However, this angle is small ($\Delta PA \sim few~deg$) both for ESO428-G014 \citep{Riffel06} and for IC 5063 \citep{Colina91}, indicating something else may contribute to this different energy dependence. Future investigations of a larger sample of CT AGNs / Seyfert 2s are needed to further probe this point.

At higher energies ($>$4 keV) the extent along the bi-cone and cross-cone direction in IC 5063 are similar, indicating symmetrical extent of the emission. Similar cases have been found in \cite{Jones21}. However, Table ~\ref{table:excess} shows that the emission along the cross-cone is less significant than that in the bi-cone.

The hard emission is likely due to reflection off molecular clouds, while the soft X-ray emission is likely from both thermal optically thin and AGN-photoionized low density ISM. The possible differences in the radial energy dependence between IC 5063 and ESO428-G014 probably then reflect differences in the molecular cloud distributions in the two galaxies.

\begin{figure}[t]
   \begin{center}
   \includegraphics[height=0.39\textheight,angle=0]{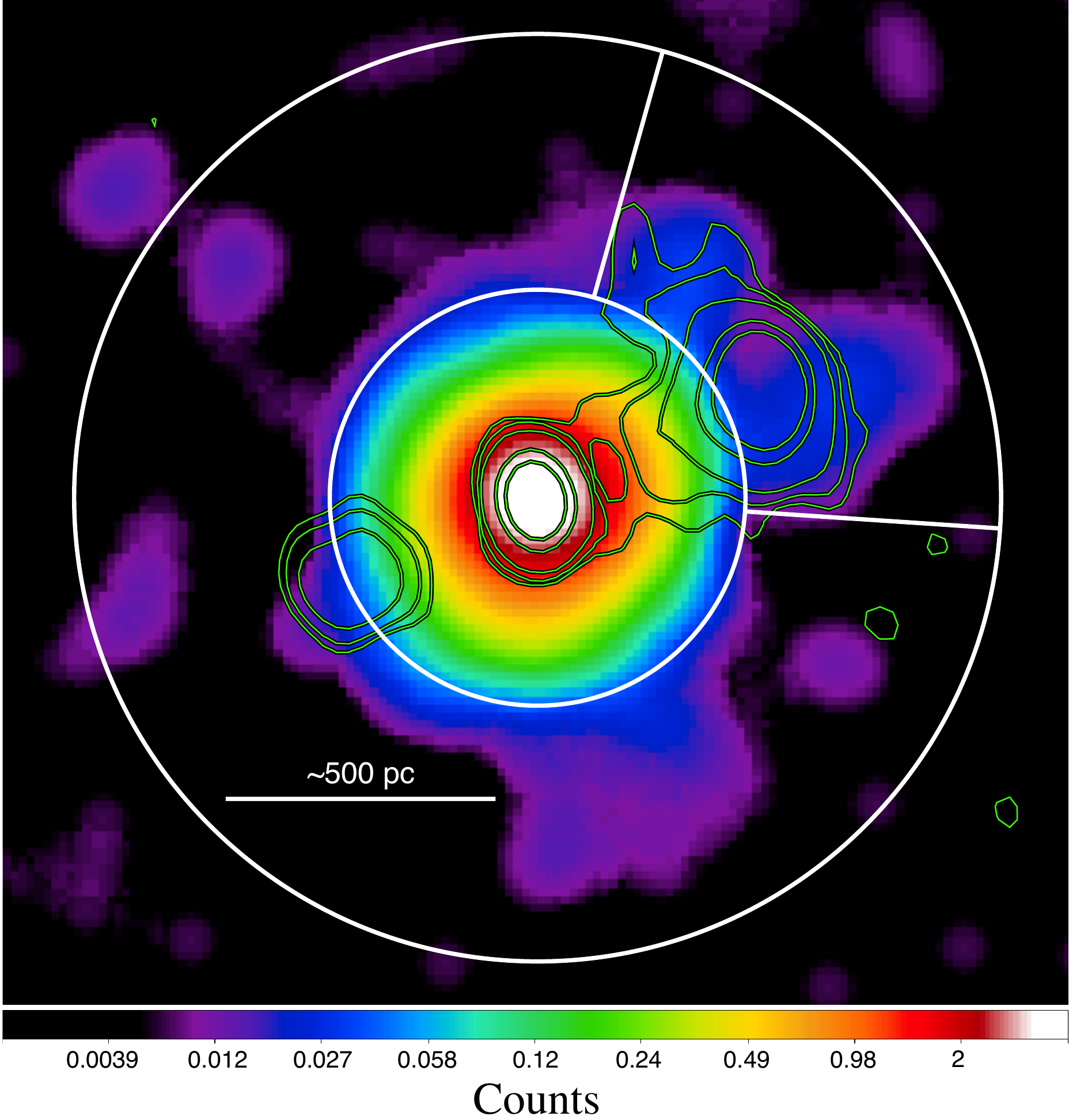}
   \caption{Merged 6.5-6.8 keV image with 1/8 of the ACIS pixel and a Gaussian smoothing with a 10 sub-pixel kernel radius, as a proxy of the ionized iron emission line morphology. In white, we report both the 1.8$''$ to 3.5$''$ annulus, and the NW cone sector in which we detect the ionized iron feature at $\sim$6.6~keV from the spectral analysis. In green we overlap radio emission contours at 17 GHz and levels 3,5,10,50,100$\sigma$ (see Fig.~\ref{image:regions}).}
   \label{image:FeIon}
   \end{center}
\end{figure}

\begin{figure}[t]
   \begin{center}
   \includegraphics[height=0.268\textheight,angle=0]{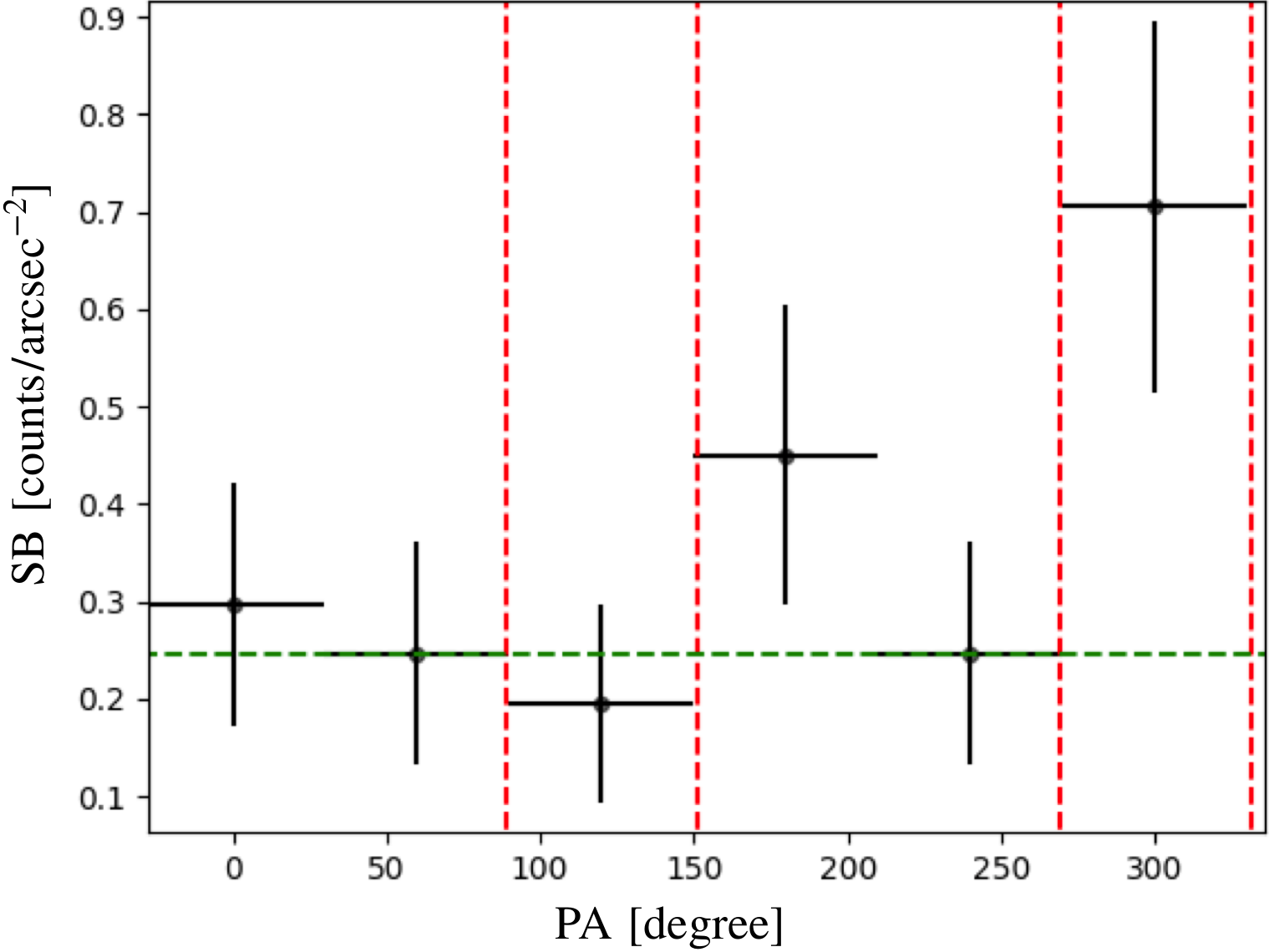}
   \caption{Surface brightness (in $\rm counts/arcesc^2$) azimuthal profile within the 432-840 pc annulus (white annulus in Fig.~\ref{image:FeIon}) in the 6.5-6.8 keV image, used as a proxy of the ionized iron line. Red dashed lines separate the bi-cone and cross-cone sectors we consider throughout the paper, while green dashed horizontal line is the average surface brightness estimated excluding the two peak values at 120$^o$ and 300$^o$.}
   \label{image:FeIon_PA}
   \end{center}
\end{figure}

\subsection{The effect of the radio jets on the hot ISM of the bi-cone} \label{sec:DiscJet}

Fig.~\ref{image:rp1} shows that the radial surface brightness profile of the bi-cone region at energies $<$1.5~keV shows a noticeable 'flattening' in the region within $\sim$3$''$ (i.e. $\sim$700~pc), a radius roughly consistent with the terminal hot spots of the radio jet in IC 5063 \citep{Morganti07}. We speculate that this enhancement of X-ray surface brightness may be connected with the interaction of the radio jet with the ISM in the same region \citep[e.g.][]{Sutherland93,Falcke98,Gallimore06,Wang09,Mukherjee18}.
The NW cone spectrum  shows a $\approx$3$\sigma$ broadband feature consistent with Fe XXV at $\sim$6.7~keV. Fig.~\ref{image:FeIon} shows a 1/8 sub-pixel image in the 6.5-6.8 keV energy band, smoothed with a kernel radius of 10 subpixel, as a proxy of the distribution of the ionized Fe XXV emission. The radio ATCA contours at 24 GHz \citep{Morganti98} are superimposed in green. This image shows a protrusion in the NW sector at a distance of 800~pc from the nucleus, which is and spatially correlated with the 24 GHz radio hot-spot. To estimate the significance of this extended ionized iron feature, we plotted the surface brightness azimuthal profile inside the 2$''$-3.5$''$ annulus (purple in Fig.~\ref{image:FeIon}) in the 6.5-6.8 keV image, with a bin size of 60 degrees. The azimuthal profile is shown in Fig.~\ref{image:FeIon_PA}, in which the dashed red lines separate the different sectors and the dashed green line represents the average value estimated in the all sectors, excluding the NW cone. We find 2.4$\sigma$ and 3.7$\sigma$ significance of the surface brightness excess in the NW sector relative to the average value and to the background, respectively. This excess is consistent with the image in Fig.~\ref{image:FeIon}. We do not detect a similar significant feature in the SW cross-cone sector.
The Fe-K$\alpha$ 6.7~keV line is typically associated with emission from collisionally (or shock) ionized gas \citep[e.g. NGC 6240]{Netzer05,Feruglio13,Wang14,Fabbiano20}. The Fe XXV line has been assumed to have a nuclear origin in most works, however, it is observed to be extended out to 40~pc in NGC 4945 \citep{Marinucci17}, and to 2.1 kpc in NGC 6240 \citep{Netzer05}, in which it appears like a bridge connecting two merged nuclei \citep{Fabbiano20}. In IC 5063 this feature is seen only in the NW cone region, where outflowing multi-phase gas has been observed \citep{Morganti07,Tadhunter14,Morganti15,Dasyra15,Dasyra16,Oosterloo17}. We speculate that its origin, as well as that of the perturbed gas, may be related to the interaction of the radio jet with the ISM.

\section{Summary and Conclusions} \label{sec:conclusion}

We have presented the spatial and spectral analyses of deep (270 ks) X-ray Chandra observations of the Seyfert 2 IC 5063. One of the most powerful radio Seyfert 2 galaxies in the local Universe ($P_{1.4~ GHz} = 3 \times 10^{23} W~Hz^{-1}$), IC 5063 is characterized by radio jets interacting with the dense galactic disks, in a region partially co-spatial with the [OIII] ionized bicone \citep{Danziger81,Morganti98}. It therefore provides a unique laboratory for investigating the interaction of nuclear activity (high-energy radiation, radio jets, and possibly nuclear winds) with the ISM of a gas-rich galaxy. In summary we find:

\begin{enumerate}[label={\arabic*)}, noitemsep]

\item The nuclear AGN spectrum (extracted from a circle of 2$''$ radius) shows both a hard continuum with an Fe-K$\alpha$ line at 6.4~keV and a soft excess at energies $>2~keV$. We model the intense hard ($>2~keV$) X-ray continuum with a reflection component and a leaky absorber with covering fraction $\simeq 99.2 \pm 0.2 \%$ and column density $\simeq 3 \times 10^{23}cm^{-2}$. The soft X-ray excess is reproduced with photoionized gas with high density ($N_H \gsim 10^{23} cm^{-2}$) and relatively low ionization parameter ($logU \sim -1.6$), mixed with either a more ionized phase of the gas ($logU \sim 1.5$) or a thermal collisionally/shock component with temperature $kT\sim 1-3$~keV (all these values are in the Table~\ref{table:parNuc}, Appendix~ \ref{sec:procedure1}).

\item Consistent with previous work \citep{Bianchi06,Wang11b,Fabbiano17,Ma20,Jones21}, most of the soft ($<3~keV$) X-ray emission is extended (out to $\sim 3.5$~kpc) along the bi-cone direction (Fig.~\ref{image:rp1}), which is also the direction of the radio jets. This extended emission is modelled with a phase of low ionization ($logU \sim -1.7, -2.7$) and less obscured ($N_H < 10^{22} cm^{-2}$) gas with respect to the nuclear component, plus a more ionized ($logU \sim 1.8$) phase of the gas or a collisionally excited gas with $kT\sim 1-1.3$~keV. The increase of soft X-ray emission along the jet (Fig.~\ref{image:rp1}) suggests jet-ISM interaction as a likely trigger for most of this emission.

\item As in \cite{Fabbiano17}, we detected kpc-scale diffuse emission of the hard (3-6~keV) X-ray continuum along the bi-cone direction. The spectrum is fitted with a reflection component steeper ($\Gamma \simeq 2.7$) than the nuclear one ($\Gamma \sim 1.5$). The 6.4~keV Fe-K$\alpha$ line is found both in the SE and NW cone spectrum. Moreover, we find a broad feature at 6.1-6.6~keV in the NW sector (Fig.~\ref{image:PAFeKa}), spatially correlated with the most intense radio hot-spot. In the same area, the NW cone spectrum suggests Fe XXV emission associated with the NW radio hot-spot (Figs.~\ref{image:FeIon}, \ref{image:FeIon_PA}). The Fe XXV ionized iron emission suggests shocks triggered by the radio jet-ISM interaction. The presence of neutral 6.4~keV Fe-K$\alpha$ emission in the same areas suggests reflection of the AGN photons from dense molecular clouds in the region. These clouds may be responsible for stopping the nuclear jet.

\item The emission at energies $<1.5$~keV show a significant ($\sim 30 \sigma$; table~\ref{table:excess}) extent ($\sim3~$~kpc; Fig.~\ref{image:TorusSoft}) along the cross-cone area, i.e. perpendicular to the ionization cone and radio jets. The soft X-ray excess of the spectrum extracted from this region is well reproduced with a mix of any 2 components among photoionization, collisional and shock ionization models, implying two possible scenarios:
\begin{itemize}
\item Most of this emission is due to photoionization by AGN \citep[as found in][]{Wang11b,Paggi12,Fabbiano18a}, thus suggesting a porosity of the obscuring torus that allow for a transmission fraction of $\sim 20 \%$, twice that estimated in ESO428-G014 \citep{Fabbiano18a}. In this case the spectrum is modelled with two highly ionized ($log U \sim 0.8-1.8$) components compared with typical values in literature \citep[e.g.][; see Fig.~\ref{image:U1U2}]{Paggi12}.

\item Thermal gas provides a large contribution to this emission, suggesting the presence of a hot ($kT \sim 0.3-2~keV$) phase outflows perpendicular to the jets direction with velocities $v_{shock} \sim 400-1000~km~s^{-1}$. This scenario is in agreement with both simulations of the radio jet-ISM interaction in IC 5063 \citep{Mukherjee18}, and optical observations of a low-ionization [SII] loop detected with HST \citep{Maksym20}, high velocity dispersion [OIII] and H$\alpha$ emission lines found in MUSE data \citep{Venturi21} and possible dust-displacing outflows in \cite{Maksym20a}, all predicting lateral hot-phase outflows. 

\end{itemize}

\item The electron density of the collisionally and shock ionized gas (table~\ref{table:phypar}) estimated in all the regions for a filling factor $\eta =1$, is much lower than the values typically observed in other AGNs \citep[e.g.][]{Paggi12,Fabbiano18a}. If the thermal gas in the cross-cone has values $n_e \sim 0.1-0.3~cm^{-3}$ as predicted in \cite{Mukherjee18} simulations, we estimate a $\eta < 0.01$. The same filling factor is also found by \cite{Oosterloo17} for CO molecular gas, suggesting a similar clumpiness of the hot and cold phase.

\end{enumerate}

In this paper we have investigated the large-scale morphology and the spectral properties of the X-ray emission of hot gas in IC 5063.
In our next paper (in preparation) we plan to perform a multi-wavelength analysis and comparison of the morphology of the diffuse emission, to better investigate the nature of multi-phase ISM gas under the effects of interaction with radio jets and outflows.

\acknowledgments
This work is partially supported by the Chandra grant GO9-20101X. AT and FF acknowledge support from PRIN MUR 2017 Black hole winds and the baryon life cycle of galaxies: the stone-guest at the galaxy evolution supper and from  from the European Union Horizon 2020 Research and Innovation Framework Programme under grant agreement AHEAD2020 n. 871158. We thank Stefano Bianchi, Margarita Karovska, Rafael Martinez Galarza, Raffaele D'Abrusco, Xiurui Zhao and Jingzhe Ma for useful discussions. We thank the referee for useful comments that have improved this work.

\vspace{5mm}
\facilities{Chandra(ACIS)}

\software{Cloudy \citep{Ferland98}, CIAO \citep{Fruscione06}, Sherpa \citep{Freeman01}}

\bibliography{andrea}{}
\bibliographystyle{aasjournal}

\appendix

\section{Spectral analysis of the Nuclear region} \label{sec:procedure1}

We analysed the nuclear spectrum over 0.3-8.5 keV energy band, extracted with the \texttt{specextract} task from a circular region of 2$''$ radius, which includes more than 90$\%$ of the PSF emission, as described in Section~\ref{Spec}. We found 29228 total net counts in the 0.3-7.0 keV energy band and a $L_{2-10~keV}=2 \times 10^{42}~erg~s^{-1}$.
For the analysis of the nuclear spectrum we did not use the full-array mode observation (i.e. 7878) in order to minimize pileup (see Sect.~\ref{merg2}).

\subsection{Leaky Absorber model of the nuclear spectrum} \label{sect:LeakAbs}

We first fitted the nuclear spectrum with a "leaky absorber" model applying a partial covering absorption model\footnote{$https://cxc.harvard.edu/sherpa/ahelp/load\_xspartcov.html$} to a power law (XSPOWERLAW\footnote{$https://cxc.cfa.harvard.edu/sherpa/ahelp/xspowerlaw.html$; to model the primary emission from the hot corona}): \texttt{partcov(xsphabs)*xspowerlaw}.

Following previous works on obscured AGN \citep{Levenson06,Fabbiano18a}, we added components to the model, guided by the $\gsim 2\sigma$ contiguous residuals and F-test\footnote{A model comparison test between two competing models of data set with a different number of degree of freedom, based on the best-fit statistics of each fit.} results. We first added a simple reflection (PEXRAV\footnote{$https://cxc.cfa.harvard.edu/sherpa/ahelp/xspexrav.html$; representing the reflection of the up-scattered photons onto the cold material of the accretion disk and the torus. We fix fold$\_$E=200, rel.refl=-1 and cosIncl=0.45.}) component with solar abundance, forcing it to have the same photon index of the power law. 
Then, we included the most prominent emission lines one at a time and left their energy and normalization free to vary, while fixing the width of the lines to the $Chandra$ spectral resolution ($\rm \sim 100~eV$). We then introduced additional emission lines to remove remnant contiguous significant residuals at $\gsim$2$\sigma$. 

\begin{figure}[t]
   \begin{center}
   \includegraphics[height=0.352\textheight,angle=0]{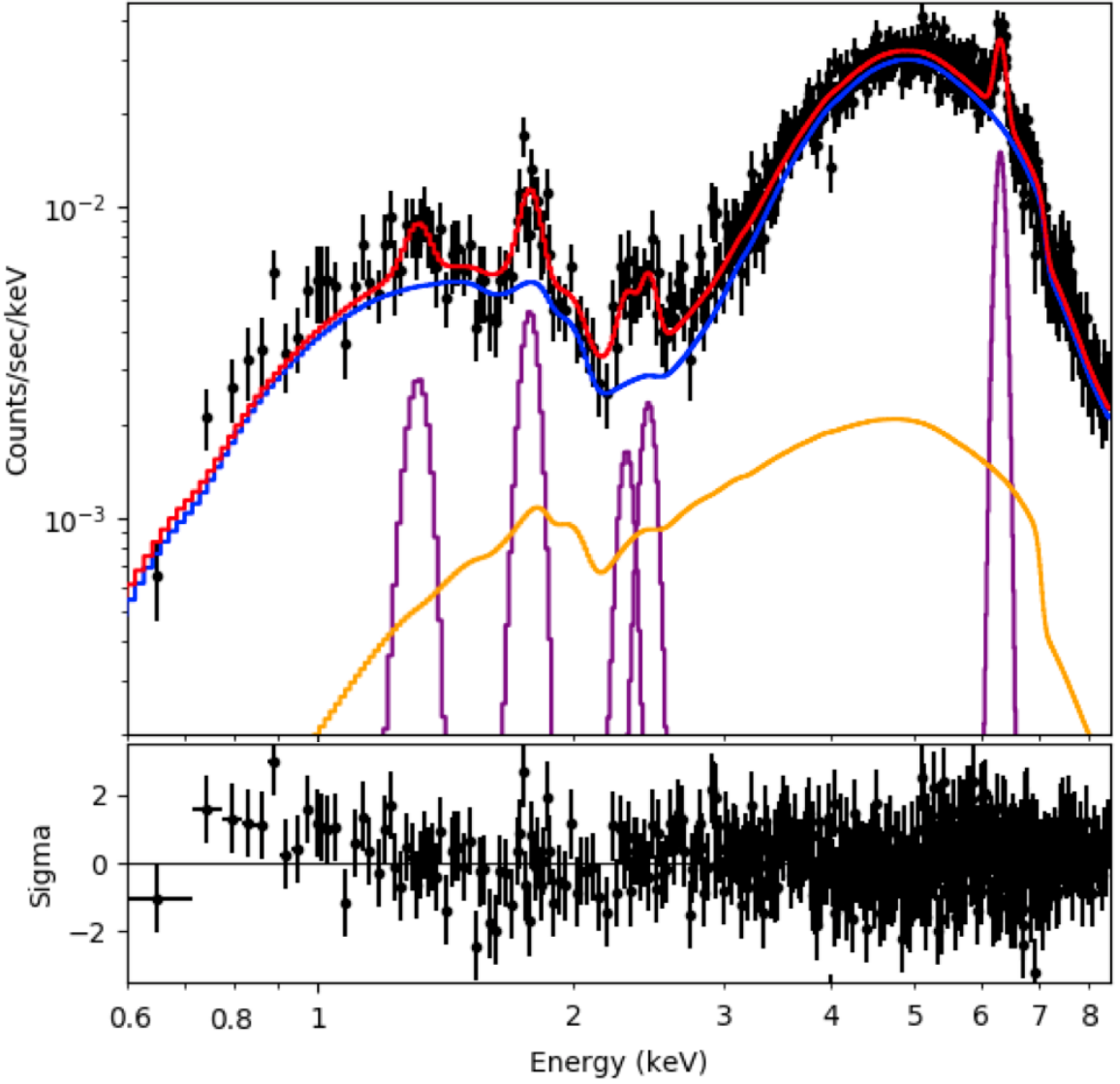}
   \caption{Leaky Absorber fit to the nuclear spectrum of IC 5063 (top panel) and residuals (bottom panel). The spectrum is extracted from a circle of 2$''$ radius ($\sim$3-4 PSF) and binned at 20 counts per bin. The total model (red line) includes a partially absorbed power law (blue line), a PEXRAV reflection component (orange line) and five Gaussian lines (purple lines).}
   \label{image:LeakyAbsorber}
   \end{center}
\end{figure}

The data, best-fit model and residuals are shown in Fig.~\ref{image:LeakyAbsorber}.  
The model yields a power law and a reflection component with photon index $\Gamma=1.45 \pm 0.10$, and $\rm N_H \simeq (2.9 \pm 0.1) \times 10^{23}~cm^{-2}$ with a covering fraction of $\rm 99.2 \pm 0.2 \%$, which is clearly required to get a good fit of the soft excess.
These results are consistent with the values observed in Seyfert 2 galaxies \citep[e.g.][]{Cappi06,Vasylenko15}. 
In addition, there are a total of five emission lines, including Fe-K$\alpha$ emission with an equivalent width (EW) $\simeq 178_{-41}^{+55}$~eV.
Table~\ref{table:parNuc} gives the best-fit parameters, and the energies and normalizations of the emission lines.

\begin{table*}[t]
\caption{Best-fit parameters and $\chi_R^2$/dof of empirical (top) and physical (bottom) best-models used to fit IC 5063 nuclear spectrum (29242 counts at 0.3-7.0 keV) in the 0.3-8.5 keV energy band. For the phenomenological models we report energy, normalization and identification of the emission lines we detect.}
\label{table:parNuc}
\resizebox{1.0\textwidth}{!}{\begin{minipage}{\textwidth}
        \begin{tabular}{c c c c c c c c}
            \toprule
\multicolumn{8}{c}{Best-fit Empirical models} \\
            \hline
\multicolumn{2}{c}{Model} & $\chi ^2$/dof & $\Gamma ~^{(a)}$  & \multicolumn{1}{c}{$\rm norm_{pl}~[ph/cm^2/s]$} & $N_H ^{pl}~[\times 10^{22} cm^{-2}]$ & \multicolumn{1}{c}{$\rm norm_{refl}~[ph/cm^2/s]$} & $N_H ^{refl}~[\times 10^{22} cm^{-2}]$ \\
             \hline 
 \multicolumn{2}{l}{(A) Leaky absorber} & 0.960/421 & $1.45 \pm 0.10$ & $2.54_{-0.44}^{+0.45} \times 10^{-3}$ & $^{(b)}$ $29.11_{-1.16}^{+1.19}$ & $1.58_{-0.55}^{+0.61} \times 10^{-3}$ & -- \\   
 \multicolumn{2}{l}{(B) 2 pl + refl} & 0.932/414 & 1.7$^{(c)}$ & $3.76_{-0.22}^{+0.18} \times 10^{-3}$ & $34.98_{-0.86}^{+0.92}$ & $7.62_{-1.41}^{+1.79} \times 10^{-3}$ & $3.11_{-0.96}^{+1.11}$ \\  
 \multicolumn{2}{l}{}    &     & 2.2 $^{(c)}$ & $2.27 \pm 0.16 \times 10^{-5}$ &    --   & -- & -- \\  
             \hline 
 \multicolumn{7}{c}{Emission lines} \\  
 & \multicolumn{1}{c}{Energy [keV]} &  \multicolumn{2}{c}{Flux [$\rm 10^{-6}~ph~cm^{-2} s^{-1}$]}& \multicolumn{4}{c}{Identified emission lines $^{(d)}$}\\   
             \hline
\multicolumn{1}{c}{(A)} &  \multicolumn{1}{c}{$1.32 \pm 0.02$} & \multicolumn{2}{c}{$0.87 \pm 0.27$} & \multicolumn{4}{c}{Mg XI  [1.331 keV]} \\
\multicolumn{1}{c}{} &  \multicolumn{1}{c}{$1.79 \pm 0.01$} & \multicolumn{2}{c}{$1.13 \pm 0.23$} & \multicolumn{4}{c}{blend Mg XII [1.745 keV] + Si XIII [1.865 keV] / Fe XXIV [1.778 keV]} \\
\multicolumn{1}{c}{} &  \multicolumn{1}{c}{$2.33 \pm 0.03$} & \multicolumn{2}{c}{$0.65 \pm 0.30$} & \multicolumn{4}{c}{Si XIII  [2.346 keV]} \\
\multicolumn{1}{c}{} &  \multicolumn{1}{c}{$2.48 \pm 0.02$} & \multicolumn{2}{c}{$0.93 \pm 0.29$} & \multicolumn{4}{c}{S XV  [2.461 keV]} \\
\multicolumn{1}{c}{} &  \multicolumn{1}{c}{$6.39 \pm 0.01$} & \multicolumn{2}{c}{$21.1 \pm 1.5$} & \multicolumn{4}{c}{Fe-K$\alpha$ [6.4 keV]} \\
\multicolumn{1}{c}{(B)} &  \multicolumn{1}{c}{$1.01 \pm 0.02$} & \multicolumn{2}{c}{$0.93 \pm 0.50$} & \multicolumn{4}{c}{Fe XXI  [1.009 keV] / Ne X [1.022 keV]} \\
\multicolumn{1}{c}{} &  \multicolumn{1}{c}{$1.33 \pm 0.01$} & \multicolumn{2}{c}{$1.28 \pm 0.28$} & \multicolumn{4}{c}{Mg XI  [1.331 keV]} \\
\multicolumn{1}{c}{} &  \multicolumn{1}{c}{$1.47 \pm 0.02$} & \multicolumn{2}{c}{$0.47 \pm 0.21$} & \multicolumn{4}{c}{Mg XII  [1.473 keV]} \\
\multicolumn{1}{c}{} &  \multicolumn{1}{c}{$1.76 \pm 0.02$} & \multicolumn{2}{c}{$1.23_{-0.40}^{+0.28}$} & \multicolumn{4}{c}{Mg XII [1.745 keV] / Fe XXIV [1.778 keV]} \\
\multicolumn{1}{c}{} &  \multicolumn{1}{c}{$1.85 \pm 0.03$} & \multicolumn{2}{c}{$0.75_{-0.31}^{+0.34}$} & \multicolumn{4}{c}{Si XIII  [1.865 keV]} \\
\multicolumn{1}{c}{} &  \multicolumn{1}{c}{$2.33 \pm 0.03$} & \multicolumn{2}{c}{$0.67 \pm 0.29$} & \multicolumn{4}{c}{Si XIII  [2.346 keV]} \\
\multicolumn{1}{c}{} &  \multicolumn{1}{c}{$2.47 \pm 0.02$} & \multicolumn{2}{c}{$0.80 \pm 0.28$} & \multicolumn{4}{c}{S XV  [2.461 keV]} \\
\multicolumn{1}{c}{} &  \multicolumn{1}{c}{$6.39 \pm 0.01$} & \multicolumn{2}{c}{$20.2 \pm 1.9$} &  \multicolumn{4}{c}{Fe-K$\alpha$ [6.4 keV]} \\
             \toprule
\multicolumn{8}{c}{Best-fit Photoionization models} \\
             \hline   
$\chi_R^2$/dof & $\Gamma ~^{(a)}$  & $N_H ^{pl}$ [$\times 10^{22}~cm^{-2}$] & CvrFract [$\%$] & $\rm log~U$ & $\rm log~[N_H / cm^{-2}]$ & $\rm kT~[keV]$ & $\rm EM ~[10^{-6} ~ cm^{-5}]$ \\[2pt]
             \hline
0.966/427 & $1.41_{-0.15}^{+0.12}$ & $29.14_{-1.68}^{+1.49}$ & $99.7_{-0.3}^{+0.2}$ & $-1.60_{-0.08}^{+0.09}$ & $>23.5$ & & \\[3pt]
          &                        &                         &                      & $1.50_{-0.11}^{+ 0.09}$ & $22.95_{-0.15}^{+0.12}$ &  &  \\[3pt]
0.970/427 & $1.31 \pm 0.11$ & $28.26_{-1.26}^{+1.33}$ & $99.4 \pm 0.2$ & $-1.50_{-0.09}^{+0.05}$ & $>23.5$ & $^{th} 1.21_{-0.18}^{+0.13}$ & $4.98_{-1.84}^{+2.13}$ \\[3pt]
0.978/427 & $1.31 \pm 0.11$ & $28.17_{-0.50}^{+0.41}$ & $99.4 \pm 0.4$ & $-1.50_{-0.05}^{+0.05}$ & $>23.3$ & $^{sh} 2.87_{-1.76}^{+2.11}$ & $6.49_{-0.90}^{+1.73}$ \\[3pt]
             \toprule
             & & & & & & & \\
        \end{tabular}
     \end{minipage}}
     {\raggedright \textbf{Notes.} $^{(a)}$ same photon index for the power law and reflection component; \par}
     {\raggedright                 $^{(b)}$ column density associated with a partial covering model with covering fraction $99.2 \pm 0.2~\%$; \par}
     {\raggedright                 $^{(c)}$ Photon indices are fixed according to \cite{Vignali97} and \cite{Tazaki11}; \par}
     {\raggedright                 $^{(d)}$ Identification emission lines from \textit{atomdb.org} database. \par}
     {\raggedright EM is the normalization of the collisional (APEC) and shock (PSHOCK) ionization model equivalent to $\frac{10^{-14}}{4 \pi [D_A (1+z)]^2} \int n_e n_H dV$, with $D_A$ the angular distance, and $n_e$, $n_H$ the electron and hydrogen density. respectively; \par}
     {\raggedright                 $^{(sh/th)}$ temperature and normalization of the shock/thermal model. \par}
\end{table*}

\subsubsection{Consistency with previous works on IC 5063}

\begin{figure}[t]
   \begin{center}
   \includegraphics[height=0.35\textheight,angle=0]{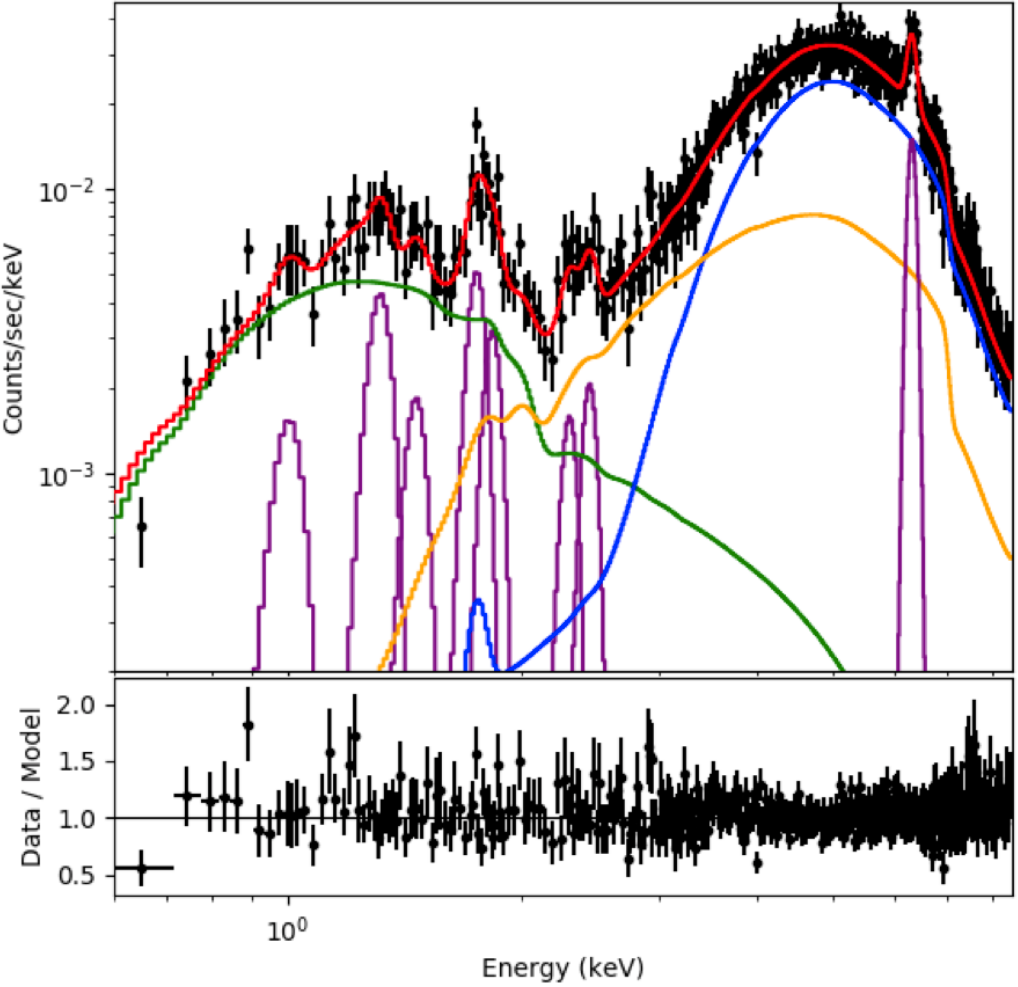}
   \caption{Soft excess power law to fit the nuclear spectrum (top panel) and residuals (bottom panel). The soft ($<$2 keV) X-ray emission is fitted with a unabsorbed power law component with $\rm \Gamma$=2.2 (green line) plus eight Gaussian emission lines. The hard ($>$2 keV) X-ray spectral fit consists of a Gaussian emission line, to model the Fe-K$\alpha$ transition, and an highly obscured power law (blue line) and an intrinsically absorbed reflection PEXRAV (orange line) with $\rm \Gamma$=1.7 and neutral hydrogen column density $\rm N_H \sim 3.5 \times 10^{23} cm^{-2}$ and $\rm N_H \sim 3.1 \times 10^{22} cm^{-2}$, respectively.}
   \label{image:fitnucleus}
   \end{center}
\end{figure} 

We estimated a total observed $L_{2-10~keV} \sim 4 \times 10^{42}~erg~s^{-1}$ similar to that of \cite{Tazaki11} ($L_{2-10~keV} \sim 5.6 \times 10^{42}~erg~s^{-1}$), both in a $\approx$2.5 arcmin radius region, but one order of magnitude lower than that of \cite{Vignali97} ($2 \times 10^{43}~erg~s^{-1}$).
We fitted the soft excess power law model used by both \cite{Vignali97} and \cite{Tazaki11} to the nuclear spectrum of IC 5063. 
This model consists of a power law plus a 6.4 keV Gaussian emission line, i.e. the neutral iron K$\alpha$ emission, and a reflection plus a fixed power law component with a photon index $\Gamma$=1.7, and an associated intrinsic absorption \citep[see][]{Tazaki11}.
For the Compton reflection component PEXRAV we fix solar abundance and the cosine of the inclination angle of the scattering disk at $cos(Incl)$=0.45.
The soft ($<$2 keV) X-ray excess is modeled with an unabsorbed $\Gamma=2.2$ power law, as often found in the spectra of Seyfert 2 galaxies \citep[e.g.][]{Guainazzi07, Bianchi09}.

Fig.~\ref{image:fitnucleus} shows the data, best-fit model and residuals, Table~\ref{table:parNuc} lists the best-fit parameters.
By fixing the photon indices as used in \cite{Vignali97} and \cite{Tazaki11}, we find a similar column density for the power law ($3.50 \pm 0.09 \times 10^{23} cm^{-2}$) and reflection ($3.11_{-0.96}^{+1.11} \times 10^{22} cm^{-2}$) components. This suggests that the shape of the nuclear spectrum of IC 5063 has remained unchanged from 1994 \citep{Vignali97} to 2009 \citep{Tazaki11} to the present.

In conclusion, we use the leaky absorber as the best-fit model, as it is the simpler one and has only marginally worse $\chi^2$ (Table~\ref{table:parNuc}).

\subsection{Physical models of the nuclear spectrum}\label{sec:phmodNuc}

\begin{figure}[t]
   \begin{center}
   \includegraphics[height=0.24\textheight,angle=0]{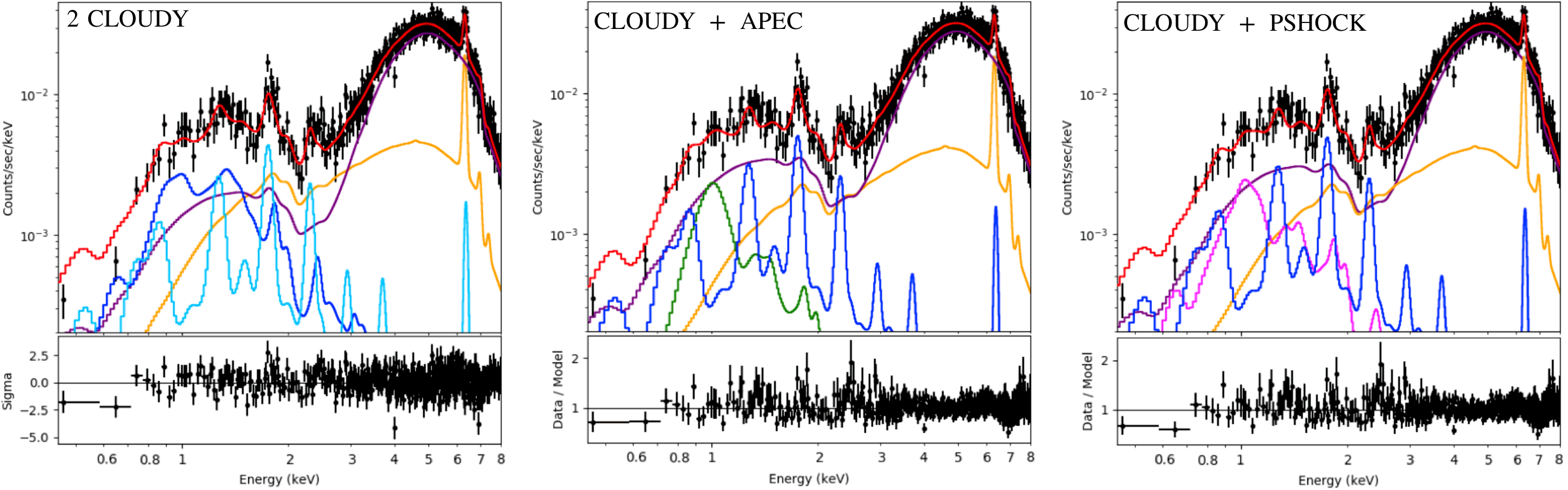}
   \caption{Best-fit physical models and residuals for the nuclear spectrum. Spectral fit consists of a leaky absorber model (purple), a PEXMON reflection component (orange), plus a mix of photoionization CLOUDY (blue and sky-blue), collisional plasma APEC (green) and shock PSHOCK (magenta) models.}
   \label{imm:phthnucleus}
   \end{center}
\end{figure}

\begin{figure}[t]
   \begin{center}
   \includegraphics[height=0.15\textheight,angle=0]{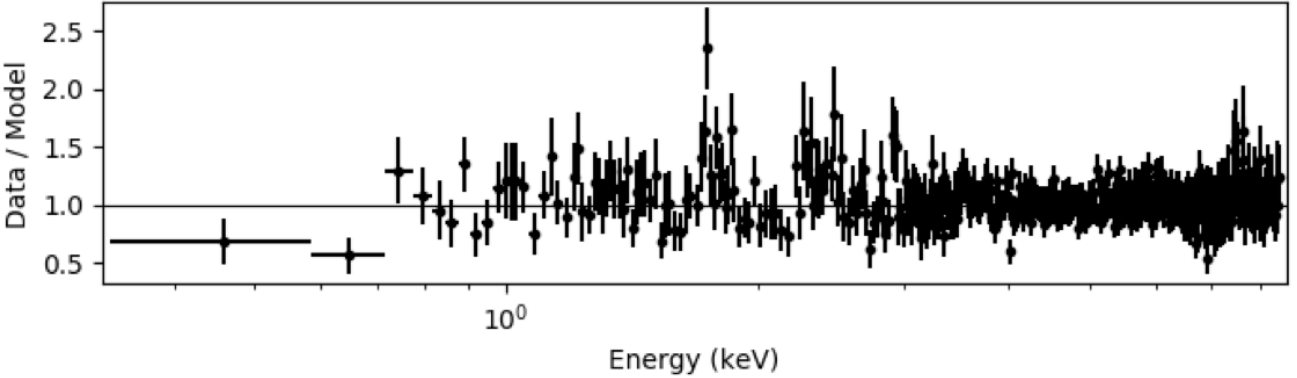}
   \caption{Residuals (data/model) obtained by fitting the nuclear spectrum with a single CLOUDY model.}
   \label{image:ResCNW0}
   \end{center}
\end{figure}

We investigated the physical mechanisms responsible for the X-ray emission in the nucleus, by fitting the emission lines in the spectrum with a combination of photoionization \citep[CLOUDY][]{Ferland98}, optically thin thermal \citep[APEC][]{Foster12}\footnote{$https://heasarc.gsfc.nasa.gov/xanadu/xspec/manual/XSmodelApec.html$} and shock \citep[PSHOCK][]{Borkowski01}\footnote{$https://heasarc.gsfc.nasa.gov/xanadu/xspec/manual/XSmodelPshock.html$} models. We followed a similar procedure to that used for the spectral analysis of ESO428-G014 \citep{Fabbiano18a}.
In particular, the photoionization model consists of a grid of values produced with the CLOUDY c08.01 package. The variables in CLOUDY are the ionization parameter\footnote{$\rm U \simeq \int _{\nu _R} ^{+ \infty} L_{\nu} d \nu / 4 \pi r^2 c n_e$ with r the distance of the gas from the source, $\rm L_{\nu}$ the ionizing luminosity, $\rm \nu _R$ the Rydberg frequency, and $\rm n_e$ the electron density.} (log~U=[-3.00:2.00] in steps of 0.25) and hydrogen column density (log $\rm N_H$=[19.5:23.5] in steps of 0.1) through the irradiated slab of gas. 
The APEC model generates a spectral emission from a collisionally-ionized, and optically thin, diffuse hot ($10^4 < T_e < 10^9~K$) plasma assuming a thermal collisional ionization equilibrium and that the collisional excitation dominates. The collisional excited plasma may be powered by a shock-confined outflow \citep[see][]{Maksym19}. 
We also considered PSHOCK model, as it allows for modelling a total or partial collisionless heating of electron in the shock front \citep{Borkowski01}. In particular, \cite{Borkowski01} assume a plane-parallel shocked plasma model with constant post-shock electron and ions temperature, element abundances, and ionization timescale, providing a useful approximation for supernova remnants, but more generally for all cases in which X-ray emission is produces in a shock front.

We added the physical models to those of the leaky absorber plus reflection models (Sect.~\ref{sect:LeakAbs}). The PEXRAV model was replaced by the reflection PEXMON\footnote{$https://cxc.cfa.harvard.edu/sherpa/ahelp/xspexmon.html$} model \citep{Nandra07}, which self-consistently generates iron and nickel emission lines. 

We initially set the normalizations of the partially absorbed power law and reflection components to zero to allow the inclusion of new models in the soft ($<$3 keV) band.
After fitting the new models all the parameters of the leaky absorber model were allowed to vary.
The choice of the best fit is based on: the statistical F-test, the shape of the residuals as the ratio between data and model, and on the physical plausibility of the fit parameters, as discussed below.

We started considering the physical models individually, and then increased the complexity of the model by adding additional components up to a maximum of 3 physical models. We estimated the significance of the improvement due to an additional component with the F-test. 

The strong emission feature at $\sim$1.7-1.9 keV, likely arising from a blend of Mg XII and Si XIII emission lines, is a clear signature of the presence of photoionized gas: the collisional (APEC) and shock (PSHOCK) models fail to reproduce this emission feature. Therefore, we at least require one photoionization CLOUDY component in each model below.

A single CLOUDY model gives a fit with good overall statistics, i.e. reduced $\chi ^2 \approx 1$, but still leaves significant ($ \sim 4 \sigma$) residuals at low energies $<$0.7 keV, at $\sim$1.7-1.9 keV and $\sim$2.3~keV (see Fig.~\ref{image:ResCNW0}).

To minimize these residuals, we fitted the spectrum with the following two-component combinations: 2 photoionization, photoionization + collisional ionization and photoionization + shocked ionization models (Fig~\ref{imm:phthnucleus}).
Based on the F-test, we obtain a significant improvement in all cases, leaving only $\sim 2.5 \sigma$ significance contiguous residuals at $< 0.7$ keV and at $\sim$1.7-1.9 keV and $\sim$2.3 keV, where the strongest emission lines are located. The best-fits are obtained by considering 2 CLOUDY components ($\chi_R ^2 /dof =0.966/427$). Adding a APEC or PSHOCK component gives similar reduced $\chi_R^2 /dof$ ($0.970/427$ and $0.978/427$ respectively). We find no improvement with three component models.
The best-fit parameters are reported in Table~\ref{table:parNuc}. The best-fit model indicates the presence of both low ($log~U =-1.6 \pm 0.1$) and high ($log~U =1.5 \pm 0.1$) photoionization gas with high column density ($log~[N_H/cm^2] >  22.9$) consistent with the high column density ($log~[N_H/cm^2] = 23.46_{-0.02}^{+0.03}$) of the direct power law in the same fit.

In the other two cases the fits suggest a low photoionization, high-density gas and either collisional emission with temperature kT$= 1.3 \pm 0.2$ keV or shocked gas with kT$= 2.87_{-1.76}^{+2.11}$~keV.

\begin{figure*}[t]
   \begin{center}
   \includegraphics[height=0.43\textheight,angle=0]{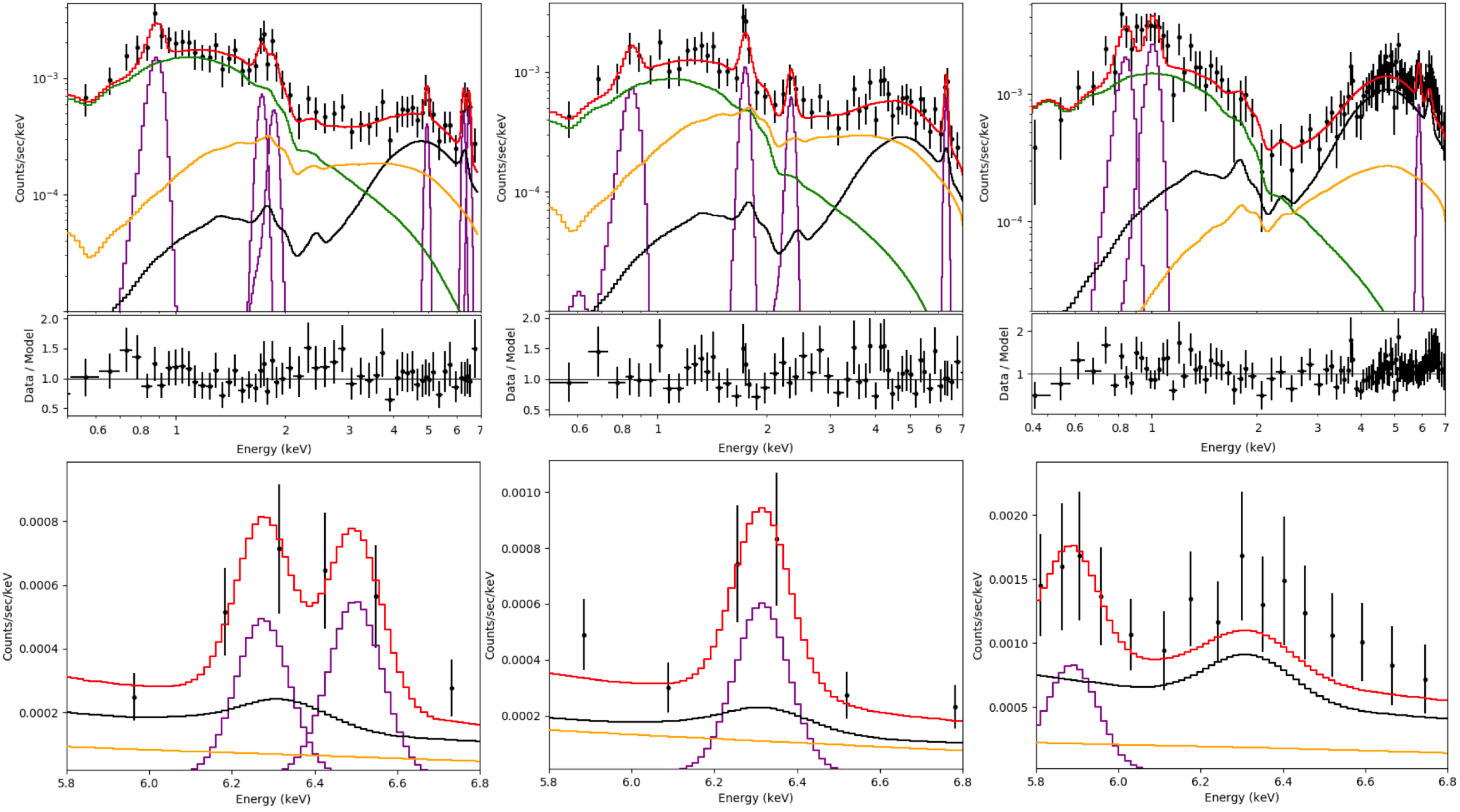}
   \caption{Top panels: spectra extracted from the NW cone (left), SE cone (center) and cross-cone (right) sectors and binned at 20 counts per bin. For each we report the best-fit empirical model and residuals. Models consists of a power law component (green), a Compton reflection PEXRAV component (orange) and some emission lines (purple). The black lines are fixed templates representing a spectral model due to the nuclear spillover plus X-ray binaries emission we derive in Sect.~\ref{sec:NucSpillover}. Bottom panels: Zoom-in of the iron transitions in the spectral region 5.8-6.8 keV of the spectra above.}
   \label{image:fitExtension}
   \end{center}
\end{figure*}

\begin{table*}[t] 
\footnotesize{
\caption{Best-fit parameters and reduced $\chi_R^2$/dof of empirical (top) and physical (bottom) models of the 0.3-7.0 keV spectra extracted from extended regions. For the phenomenological models we report energy, normalization and identification of the emission lines we detect.}
\label{table:parExtReg}
\resizebox{0.9\textwidth}{!}{\begin{minipage}{\textwidth}
        \begin{tabular}{c c c c c c c c}
            \toprule
            \hline
Regions & $\chi ^2$/dof & \multicolumn{1}{c}{counts (0.3-7.0 keV)} & \multicolumn{1}{c}{$\Gamma~^{(a)}$}  & \multicolumn{2}{c}{$\rm norm_{pl} ~[\times 10^{-6}~ph/cm^2/s]$}  & \multicolumn{1}{c}{$\rm norm_{refl}~[\times 10^{-4}~ph/cm^2/s]$} & \multicolumn{1}{c}{$\rm L_{2-10~keV}~[10^{40}~erg~s^{-1}]$} \\
             \hline    
 \multicolumn{1}{l}{(A) NW cone} & 0.600/44 & \multicolumn{1}{c}{1332} & \multicolumn{1}{c}{$2.70_{-0.28}^{+0.30}$} & \multicolumn{2}{c}{$7.10_{-0.92}^{+0.69}$} & \multicolumn{1}{c}{$6.18_{-3.14}^{5.25}$} & \multicolumn{1}{c}{$2.5$} \\  
 \multicolumn{1}{l}{(B) SE cone} & 0.700/45 & \multicolumn{1}{c}{1220} & \multicolumn{1}{c}{$2.70_{-0.44}^{+0.40}$} & \multicolumn{2}{c}{$4.12_{-1.18}^{+0.79}$} & \multicolumn{1}{c}{$9.79_{-5.42}^{+8.37}$}  & \multicolumn{1}{c}{$2.5$} \\
 \multicolumn{1}{l}{(C) Cross-cone} & 0.736/91 & \multicolumn{1}{c}{2283} & \multicolumn{1}{c}{$3.15 \pm 0.20$; $<1.2$ (refl)} & \multicolumn{2}{c}{$6.97 \pm 0.69$} & \multicolumn{1}{c}{$1.05 \pm 0.20$}   & \multicolumn{1}{c}{$6.3$}\\
             \hline  
 \multicolumn{7}{c}{Emission lines} \\ 
 & \multicolumn{1}{c}{Energy [keV]} &  \multicolumn{2}{c}{Flux [$\rm 10^{-6}~ph~cm^{-2} s^{-1}$]}& \multicolumn{4}{c}{Identified emission lines $^{(b)}$}\\   
             \hline
\multicolumn{1}{c}{(A)}   &  \multicolumn{1}{c}{$0.89 \pm 0.02$} & \multicolumn{2}{c}{$1.18 \pm 0.40$} & \multicolumn{4}{c}{Ne IX [0.905 keV] / Fe XVII [0.897 keV]} \\
\multicolumn{1}{c}{}   &  \multicolumn{1}{c}{$1.74 \pm 0.03$} & \multicolumn{2}{c}{$0.17 \pm 0.10$} & \multicolumn{4}{c}{Mg XII [1.745 keV]} \\
\multicolumn{1}{c}{}   &  \multicolumn{1}{c}{$1.90 \pm 0.05$} & \multicolumn{2}{c}{$0.15 \pm 0.10$} & \multicolumn{4}{c}{Si XIII [1.865 keV]} \\
\multicolumn{1}{c}{}   &  \multicolumn{1}{c}{$5.02_{-0.05}^{+0.06}$} & \multicolumn{2}{c}{$0.19 \pm 0.11$} & \multicolumn{4}{c}{Ti XXII [4.977 keV]} \\
\multicolumn{1}{c}{}   &  \multicolumn{1}{c}{$6.34_{-0.04}^{+0.34}$} & \multicolumn{2}{c}{$0.48 \pm 0.20$} & \multicolumn{4}{c}{Fe K$\alpha$ [6.4038 keV]} \\
\multicolumn{1}{c}{}   &  \multicolumn{1}{c}{$6.57_{-0.04}^{+0.12}$} & \multicolumn{2}{c}{$0.61 \pm 0.23$} & \multicolumn{4}{c}{Fe Be-, Li-like K$\alpha$ [6.629,6.653 keV]/Fe XXV [6.610 keV]} \\[1.5pt]
\multicolumn{1}{c}{(B)}   &  \multicolumn{1}{c}{$0.89 \pm 0.02$} & \multicolumn{2}{c}{$0.67 \pm 0.38$} & \multicolumn{4}{c}{Ne IX [0.905 keV] / Fe XVII [0.897 keV]} \\
\multicolumn{1}{c}{}   &  \multicolumn{1}{c}{$1.76_{-0.01}^{+0.02}$} & \multicolumn{2}{c}{$0.25 \pm 0.10$} & \multicolumn{4}{c}{Mg XII [1.745 keV]} \\
\multicolumn{1}{c}{}   &  \multicolumn{1}{c}{$2.37_{-0.06}^{+0.04}$} & \multicolumn{2}{c}{$0.25 \pm 0.11$} & \multicolumn{4}{c}{Si XIV [2.377 keV]} \\
\multicolumn{1}{c}{}   &  \multicolumn{1}{c}{$6.38 \pm 0.03$} & \multicolumn{2}{c}{$0.59 \pm 0.21$} & \multicolumn{4}{c}{Fe K$\alpha$ [6.4038 keV]} \\
\multicolumn{1}{c}{(C)}   &  \multicolumn{1}{c}{$0.86 \pm 0.03$} & \multicolumn{2}{c}{$1.87 \pm 0.58$} & \multicolumn{4}{c}{Ne IX [0.905 keV] / Fe XVII [0.897 keV]} \\
\multicolumn{1}{c}{}   &  \multicolumn{1}{c}{$1.02 \pm 0.02$} & \multicolumn{2}{c}{$1.25\pm 0.31$} & \multicolumn{4}{c}{Fe XXI [1.009 keV] / Ne X [1.022 keV]} \\
\multicolumn{1}{c}{}   &  \multicolumn{1}{c}{$5.95 \pm 0.04$} & \multicolumn{2}{c}{$0.64 \pm 0.24$} & \multicolumn{4}{c}{Cr XXIV [5.932 keV]} \\
             \toprule
\multicolumn{8}{c}{Best-fit Physical models} \\
             \hline   
Spectrum$^{(c)}$ & \multicolumn{1}{c}{Fit-Models} & $\chi_R^2$/dof & $\Gamma_{refl}$ & $\rm log~U$ & $\rm log~(N_H / cm^{-2})$ & $\rm kT~[keV]$ & $\rm EM ~[10^{-6} ~ cm^{-5}]$ \\[2pt]
             \hline
\multicolumn{1}{c|}{NW cone} & \multicolumn{1}{c}{2 CLOUDY}  & 0.647/53 & $2.14_{-0.53}^{+0.79}$ & $1.88_{-0.11}^{+0.08}$ & $<19.94$ & \\[2pt]
\multicolumn{1}{c|}{}  & \multicolumn{1}{c}{} &          &  & $-2.77$ $^{(d)}$ & $<20.73$ & \\[2pt]
\multicolumn{1}{c|}{} & \multicolumn{1}{c}{CLOUDY + APEC} &  0.674/53 & $2.10_{-0.52}^{+0.32}$ & $-2.72$ & $<20.47$ & $1.01_{-0.13}^{+0.27}$ & $1.84_{-0.45}^{+0.46}$ \\[2pt]
      \hline
\multicolumn{1}{c|}{SE cone} & \multicolumn{1}{c}{CLOUDY + APEC}  & 0.610/50 & $>$2.5 & $-1.83_{-0.12}^{+0.10}$ & $22.15_{-0.24}^{+0.25}$ & $1.36_{-0.19}^{+0.32}$ & $1.79_{-0.71}^{+1.07}$ \\[2pt]
\multicolumn{1}{c|}{}   & \multicolumn{1}{c}{2 CLOUDY} & 0.613/50 & $2.49_{-0.25}^{+0.30}$ & $1.75 ~^{(d)}$ & $21.61_{-0.24}^{+0.17}$ & & \\[2pt]
\multicolumn{1}{c|}{}  & \multicolumn{1}{c}{} &  &  & $-1.72_{-0.16}^{+0.12}$ & $22.33_{-0.27}^{+0.29}$ & &   \\[2pt]
 \hline
\multicolumn{1}{c|}{Cross-cone} & \multicolumn{1}{c}{2 PSHOCK} & $0.750/35$ & $<1.1$ & & & $^{(sh)}$ $0.68_{-0.19}^{+0.12}$ & $28.5_{-9.7}^{+6.4}$ \\[2pt]
\multicolumn{1}{c|}{} & \multicolumn{1}{c}{}    &       &  & & & $^{(sh)}$ $1.03_{-0.11}^{+0.16}$ & $3.41_{-0.56}^{+0.32}$ \\[2pt]
\multicolumn{1}{c|}{} & \multicolumn{1}{c}{CLOUDY + PSHOCK} & $0.786/35$ & $<1.1$ & $0.76_{-0.16}^{+0.20}$ & $<19.59$ & $^{(sh)}$ $1.71_{-0.34}^{+0.53}$ & $2.73_{-0.82}^{+0.79}$ \\[2pt]
\multicolumn{1}{c|}{} & \multicolumn{1}{c}{PSHOCK + APEC} & $0.755/35$ & $<1.1$ & & & $^{(sh)}$ $2.17_{-0.43}^{+0.88}$ & $3.14_{-0.47}^{+0.45}$ \\[2pt]
\multicolumn{1}{c|}{} & \multicolumn{1}{c}{} &            &  & & & $^{(th)}$ $0.54_{-0.20}^{+0.15}$ & $1.95_{-0.79}^{+1.23}$ \\[2pt]
\multicolumn{1}{c|}{} & \multicolumn{1}{c}{2 CLOUDY} & $0.788/35$ & $<1.1$ & $1.75_{-0.09}^{+0.07}$ & $<20.6$ &  &  \\[2pt]
\multicolumn{1}{c|}{} & \multicolumn{1}{c}{} &            &  & $0.75_{-0.04}^{+0.02}$ & $<19.7$ &  &  \\[2pt]
\multicolumn{1}{c|}{} & \multicolumn{1}{c}{CLOUDY + APEC} & $0.809/35$ & $<1.1$ & $0.75_{-0.08}^{+0.05}$ & $<19.6$ & $^{(th)}$ $1.23_{-0.15}^{+0.13}$ & $3.51_{-1.03}^{+0.91}$ \\[2pt]
\multicolumn{1}{c|}{} & \multicolumn{1}{c}{2 APEC} & $0.848/35$ & $<1.1$ & & & $^{(th)}$ $1.21_{-0.09}^{+0.08}$ & $5.69_{-0.80}^{+0.80}$ \\[2pt]
\multicolumn{1}{c|}{} & \multicolumn{1}{c}{} &            &  & & & $^{(th)}$ $0.30_{-0.05}^{+0.08}$ & $6.03_{-2.58}^{+2.50}$ \\[2pt]
             \toprule
             & & & & & & & \\
        \end{tabular}
     \end{minipage}}

     {\raggedright  \textbf{Notes.} We report the $\chi_R^2$/dof of the physical models to the cross-cone spectrum estimated in the 0.3-3.0 keV energy band. The rest of the statistic is evaluated in the 0.3-7.0 keV energy band. \par}
     {\raggedright                 $^{(a)}$ same photon index for the power law and reflection component, but for the cross-cone spectral model; \par}
     {\raggedright                 $^{(b)}$ Identification emission lines from \textit{atomdb.org} database; \par}
     {\raggedright                 $^{(c)}$ for each spectrum we find more than one best-fit physical model; \par}
     {\raggedright                 $^{(d)}$ parameter fixed to the best-fit value or not constrained. \par}
     {\raggedright                 $^{(sh/th)}$ temperature and normalization of the shock/thermal model. \par}
}
\end{table*}

\section{Spectral analysis of the extended regions}\label{sec:procedure2} 

Here we report the results of the spectral analysis of the diffuse gas from 2$''$ to 15$''$ (0.5-3.6 kpc), in the NW, SE and cross-cone sectors (see Sect.~\ref{Spec}). We are interested exclusively in exploring the extended emission. For this purpose, the PSF wing contribution of the nuclear emission has to be excluded. We therefore modeled the nuclear spillover spectral component for each region, described in detail in Sect~\ref{sec:NucSpillover}. We also estimated the X-ray binary (XRB) contribution to the X-ray emission at 2-10 keV and found it to be negligible ($< 2.2 \%$ with respect the total X-ray 2-10 keV emission) in all cases.

\subsection{Nuclear spill over and X-ray binaries contribution}\label{sec:NucSpillover}

\begin{figure}[t]
   \begin{center}
   \includegraphics[height=0.3\textheight,angle=0]{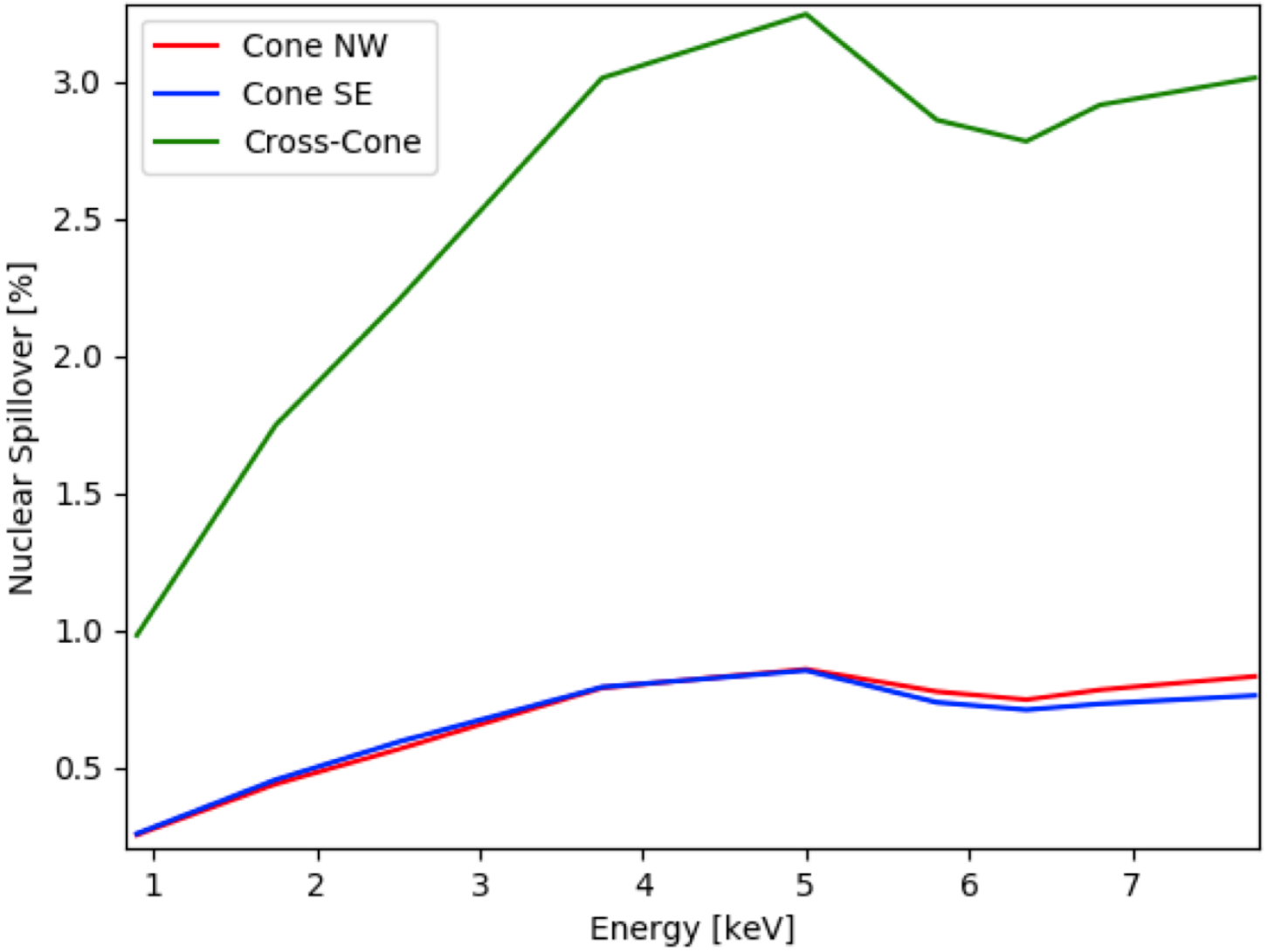}
   \caption{Contribution (in $\%$) of the nuclear spillover in the different extended regions as a function of the energy.}
   \label{image:NSO}
   \end{center}
\end{figure}

To model the spectra of the extended regions, we need first to evaluate and remove the contamination of the strong nuclear spectrum, spilling outside the central 2$''$ region in each region, due to the PSF wings\footnote{$https://cxc.harvard.edu/ciao/ahelp/psf.html$}. The contribution of the nuclear emission in the external regions is estimated from the PSF model of the merged image. This PSF model was obtained by combining 500 PSF realisations produced for each observation with the \texttt{ChaRT} and \texttt{MARX 5.5.0} tools, to give the same signal-to-noise as the observation. As $Chandra$ PSF change with energy, we produced multiple PSF models at different energy bands. 
To evaluate the residual emission due to the nucleus in each sector we computed, in each PSF image, the ratio between the net counts in a given external region and in the central circle of 2$''$ radius. These ratios represent the percentage of nuclear emission contributing to the spectral emission in each external region, and are used to rescale the best-fit of the nuclear spectrum thus producing the final nuclear spillover models.

Fig.~\ref{image:NSO} shows the percentage of nuclear emission contribution in each region versus energy. Notice that, for the single cross-cone and bi-cone sectors the PSF spillover is approximately equal to and below 1$\%$. For the entire annular extended region the fraction of nuclear spill over reaches a peak of 5$\%$ at 5 keV.
However, the PSF models are subject to statistical uncertainties estimated to be less than 10$\%$ within 2$''$ of the centroid for on-axis PSF, with a additional uncertainty $\sim$5$\%$ for off-axis PSF\footnote{$https://cxc.harvard.edu/cal/docs/cal\_present\_status.html$}. 
We therefore assume that errors of our nuclear spillover are $\sim$10-20$\%$.

Another contamination to the X-ray spectra of the extended regions, although less significant than nuclear spillover, derives from the emission of the stellar population. However, while this X-ray stellar contribution is negligible with respect to the nuclear spectrum, it may be of greater relevance in regions further away from the AGN. In particular the X-ray binaries (XRBs) are the main contributors in the 2-10 keV band \citep{Persic02}.

We used the SFR-$\rm L_{2-10~keV}$ correlation in \cite{Lehmer10} to estimate the total X-ray emission expected from the XRBs population in IC 5063. The SFR was obtained from the $\rm L_{FIR}$-SFR relation in \cite{Satyapal05}, with a far-infrared luminosity within a $\sim$12 kpc radius region of $\rm L_{FIR}=4.7 \times 10^{10}~L_{\odot}$ according to \cite{Wiklind95,Morganti98}. From these calculations, we predicted a total luminosity from the XBs of $\rm L_{2-10~keV} = 2.5_{-1.1}^{+5.9}\times 10^{40} ~erg~s^{-1}$.
Conservatively, assuming a uniform distribution of the XRBs population, we derived the XRBs luminosity at 2-10 keV expected in the extended regions, depending on their area relative to the total area of a circle of 12 kpc radius (where the $\rm L_{FIR}$ was extracted). 
Based on \citep{Persic02}, a power law with a fixed photon index $\Gamma = 1.2$ as spectral shape for the XRBs emission was assumed.
In conclusion, in each region the XRBs emission at the 2-10 keV energy band is $\approx$3.5$\%$ of the emission due to the nuclear spill over in the same energy band, i.e. 0.1$\%$ of the central 2$''$ counts. The XRB contribution is never more than 2.2$\%$ of the observed counts into NW, SE and cross-cone regions, and so can be safely neglected.

\subsection{Empirical fits}\label{sec:extfitsph}

We fitted the spectra extracted from the outer regions (i.e. NW cone, SE cone and cross-cone) with phenomenological models plus emission lines. For each spectrum we fixed the nuclear spillover component (Section~\ref{sec:NucSpillover}).

We first fitted a power law soft excess model as in Section~\ref{sect:LeakAbs}. 
Fitting with only power law plus prominent lines, we obtain a $\chi ^2 \approx 1$ and few significant residuals at $< 0.8$~keV in the bi-cone and $> 3$~keV  energies in the cross-cone spectra, respectively. The residuals were removed by adding a reflection PEXRAV component with photon index linked to the power law, significantly improving the statistics ($\chi ^2 \leq 0.9$) according to the F-test.

In the bi-cone spectra the reflection component is required to fit the hard ($>$3~keV) X-ray part of the spectrum with a roughly equal contribution from the nuclear spillover component. Instead, in the cross-cone spectrum the nuclear spillover contributes $\sim$80$\% \pm$~20$\%$ (Section~\ref{sec:NucSpillover}) of the hard X-ray continuum. 
It could explain the totality of the emission within uncertainties, so that a reflection component would be not required.
However, as the nuclear spillover model could change differently at different energies, we decided to keep its amplitude fixed, and to add a reflection PEXRAV component with photon index free to vary, in order to remove contiguous residuals at high energy ($>$5~keV).

In summary, the best-fit models (top panels; Fig.~\ref{image:fitExtension}) of all spectra consist of a power law, a PEXRAV reflection component and Gaussian emission lines. The best-fit properties and the identified emission lines are reported in Table~\ref{table:parExtReg}, and we also show the $L_{2-10~keV}$ estimated in each region. \\

The bottom panels in Fig.~\ref{image:fitExtension} show the 5.8-6.8 keV energy band, which includes the neutral and ionized Fe K emission lines. In the bi-cone the 6.4~keV neutral iron transition is double the expected contribution of the nuclear spillover. 
In the NW and SE cone spectra we find Fe K$\alpha$ lines with EW$\simeq 898_{-564}^{+1350}$~eV and $\simeq 796_{-533}^{+1612}$~eV, respectively.
Instead, the Fe K$\alpha$ line in the cross-cone spectrum is consistent with the nuclear spillover alone.

The NW cone spectrum has a weak broad feature, that can be fitted with the Fe K$\alpha$ line plus an emission line at $6.57_{-0.04}^{+0.12}$~keV, with EW$\simeq 1216_{-612}^{+1249}$~eV at 2.7$\sigma$ significance. This energy is consistent with the Fe XXV 6.7 keV line at 1.1$\sigma$, and has roughly the same intensity as the neutral iron line.\\

\subsection{Physical Models in the extended regions} \label{sect:ExtePhysicModels}

In this section, we examine the mechanisms responsible for the X-ray emission in extended regions. As in Sect.~\ref{sec:phmodNuc} for the nuclear spectrum, we fit a combination of CLOUDY, PSHOCK and APEC (with solar abundances) models. In this case we also include the nuclear spillover contribution (Section~\ref{sec:NucSpillover}).
We initially fit the spectra with a single physical component, and then added additional components as required. 
We added up to four components to the models as in Section~\ref{sec:phmodNuc}.

We show the Data/Model residuals, $\chi_R^2$ and degree of freedom of the intermediate spectral fits in Fig.~\ref{image:ResCNW}, while images of the selected best-fit models, with residuals, are reported in the following sections.

\begin{figure}[t]
   \begin{center}
   \includegraphics[height=0.28\textheight,angle=0]{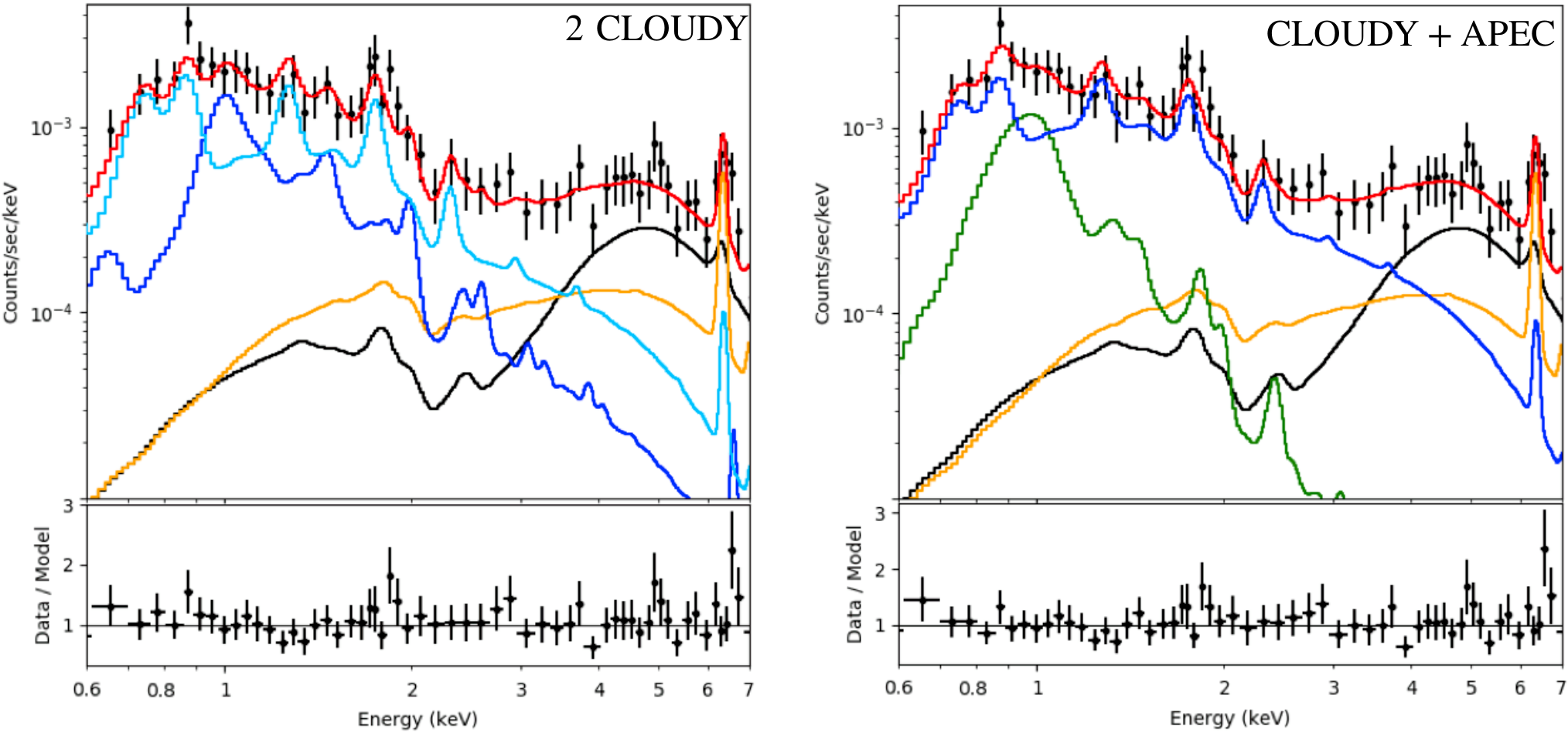}
   \caption{Spectra extracted in the NW sector, best-fit physical models and Data/Model residuals. Models consists of CLOUDY (blue, sky-blue lines), APEC (green, greenlime) and PEXMON (orange) components. The black line is the nuclear spillover plus XBs spectral emission.}
   \label{image:CNWPhMo}
   \end{center}
\end{figure}

\begin{figure}[t]
   \begin{center}
   \includegraphics[height=0.28\textheight,angle=0]{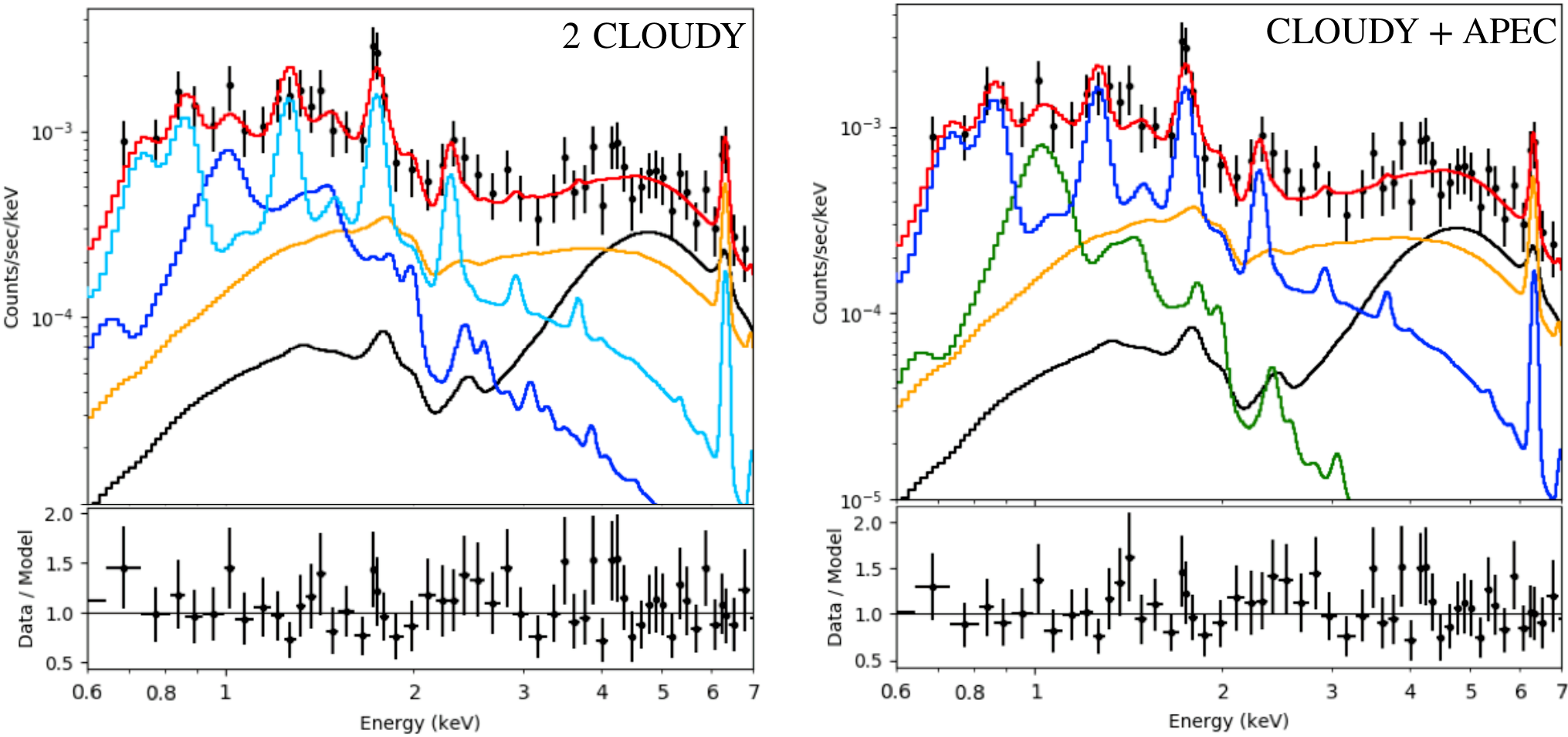}
   \caption{Spectra extracted in the SE sector, best-fit physical models and Data/Model residuals. Models consists of CLOUDY (blue, sky-blue lines), APEC (green, greenlime) and PEXMON (orange) components. The black line is the nuclear spillover plus XBs spectral emission.}
   \label{image:CSEPhMo}
   \end{center}
\end{figure}

\subsubsection{NW cone}\label{sect:ConeNWPhMod}

We first fitted the NW cone spectrum with one-component models. A single phase of collisionally or shock ionized gas fails to model the total spectrum because, as for the nuclear spectrum, these components are not able to fit the intense emission feature at $\sim$1.7-1.9 keV.
Using a single photoionization CLOUDY component we obtain a good statistic ($\chi_R^2 = 1.033$), but also some significant ($> 2\sigma$) residuals as shown in Fig.~\ref{image:ResCNW}, suggesting we need to additional model components. 

Therefore, we applied the following two-component combinations: 2 photoionization, photoionization + collisional ionization, photoionization + PEXMON reflection and photoionization + shock ionization component.
All these combinations provide a good $\chi_R ^2 \approx 1$, however they fail to model the neutral ($\simeq$6.4 keV) and ionized ($\simeq$6.6 keV) iron emission lines, as well as the emission at $\sim$1 keV and $\sim$2.3 keV.

In particular, to fit the hard ($\gsim$2~keV) X-ray part of the spectrum, we need to add a reflection PEXMON model, that gives a reduced $\chi_R ^2 \lsim$ 0.8 and no significant ($\lsim$2$\sigma$) contiguous residuals. 
Therefore, we considered the three-component combinations consisting of a PEXMON plus either 2 photoionization or photoionization + collisional ionization, or photoionization + shock ionization model. We obtain no improvement by adding further components to the 2-component + PEXMON models. 

Using PEXMON plus photoionization ($\chi_R^2$=0.674) models and plus photoionization + collisional ionization ($\chi_R^2$=0.647) models improves the $\chi_R^2$ at $\sim$30$\%$ confidence level, relative to the model including a shock ionized gas ($\chi_R^2$=0.803). 
We therefore consider the combination of photoionization and the collisionally components as the best-fit models.

Both these models require the presence of a low photoionization ($log~U \approx -2.7$) component, with column density $log[N_H/cm^2] < 20.7$, and a reflected power law with photon index $\Gamma \approx 2.1$ and normalization $\approx 1.5 \times 10^{-4} ~ph/cm^2/s$, i.e. $\sim$10 times lower than that of the nuclear reflection component. In addition, the fit suggests either a thermal $kT \simeq 1$~keV gas or a highly photo-ionized ($log~U \approx 1.88$) gas with column density $log[N_H/cm^2] < 20$. In both cases, the PEXMON template accounts for $>$80$\%$ of the 6.4~keV emission line, linking it to reflection.

\subsubsection{SE cone}

The SE cone exhibits photoionization features similar to the NW cone spectrum, except for a less intense emission peak at $\sim$1~keV with respect to the 1.7-1.9 keV emission lines. Fitting these lines requires at least one CLOUDY component of photoionized gas emission in composite models. 
All the two-component model fits yield $\chi_R^2 \leq 1$, but most exhibit significant ($\geq$2$\sigma$) contiguous residuals at $>$3 keV (Fig.~\ref{image:ResCNW}). The model including shock ionized emission is discarded due to implausible high temperatures predicted for the shocked gas ($>$10~keV).  

As in the NW cone a PEXMON reflection is included to fit the hard ($>$2$~keV$) X-ray continuum and the 6.4~keV neutral iron emission line. 
Using a photoionization + PEXMON reflection model reduces the residuals at $>$3 $keV$, achieving a reduced $\chi_R^2 = 0.983$. However residuals remain at $\approx$1.75 keV and 2.2 $keV$, suggest additional spectral components.
No significant residuals are found by using three-component models: a reflection and a photoionized phase plus, either another photoionization or a collisionally ionized component. Including a shock model creates more residuals (Fig.~\ref{image:ResCNW}).
These two best-fit models, and residuals, are shown in Fig.~\ref{image:CSEPhMo}, and the best-fit parameters are reported in Table~\ref{table:parExtReg}.
In both cases we have a low photoionization gas ($log~U < -1.7$) with $log[N_H/cm^2] \approx 22$, and a reflected PEXMON component with a photon-index $\Gamma \geq 2.5$. 
The remaining emission is ambiguously fitted either by a photoionization component with a large photoionization parameter ($log~U = 1.75$) and column density $log[N_H/cm^2] \approx 21.6$, or a collisional APEC component with temperature $kT \simeq 1.36~keV$.

\begin{figure*}[t]
   \begin{center}
   \includegraphics[height=0.495\textheight,angle=0]{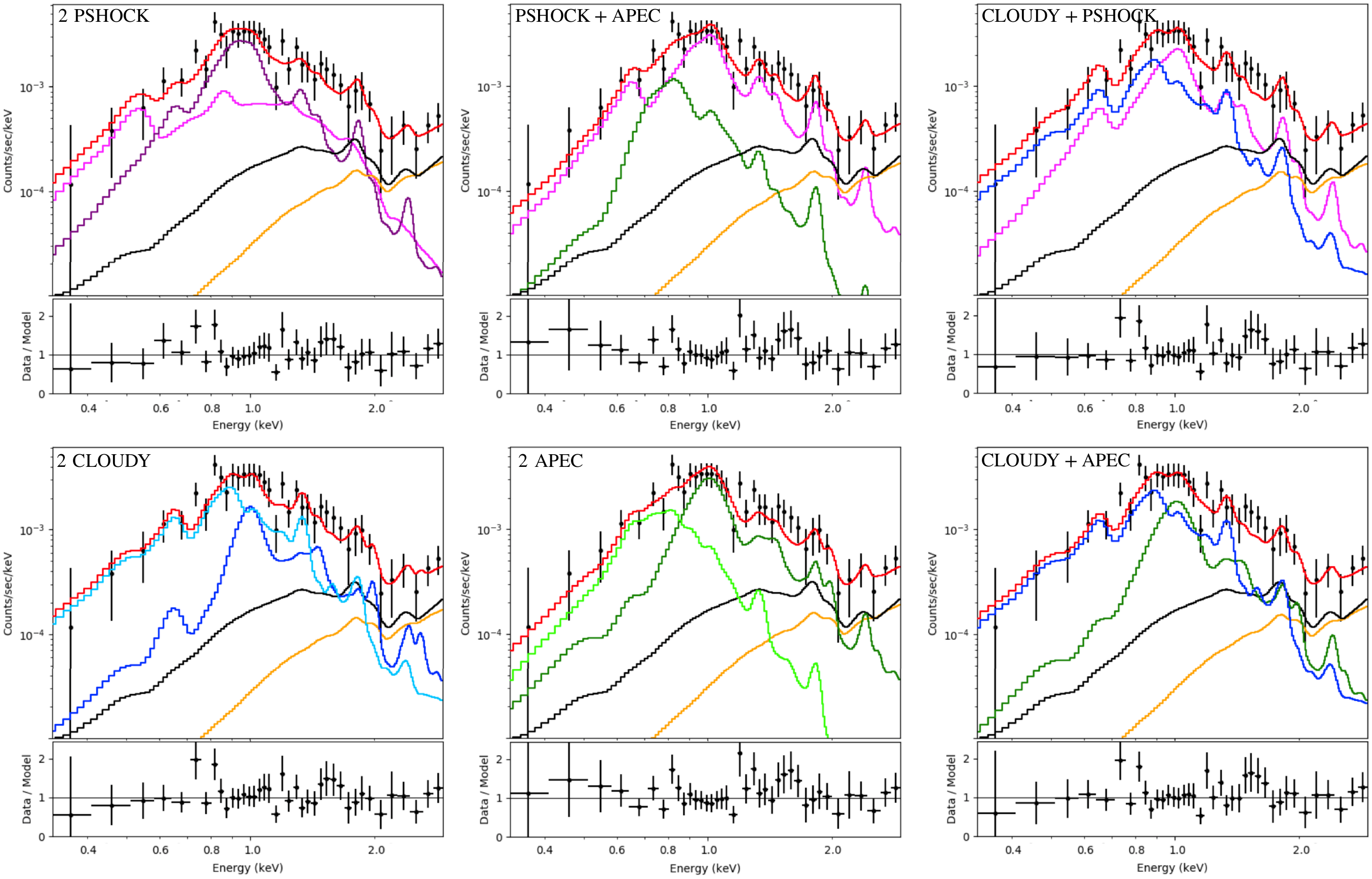}
   \caption{Soft (0.3-3 keV) X-ray spectrum extracted in the cross-cone sectors, best-fit physical models (red lines) and Data/Model residuals. Models consists of CLOUDY (blue, sky-blue lines), APEC (green, greenlime) and PSHOCK (magenta, purple lines) components. The black and orange lines are the nuclear spillover plus X-ray binaries spectral emission and the PEXRAV reflection component, respectively.}
   \label{image:CCPhMo}
   \end{center}
\end{figure*}

\begin{figure}[t]
   \begin{center}
   \includegraphics[height=0.5\textheight,angle=0]{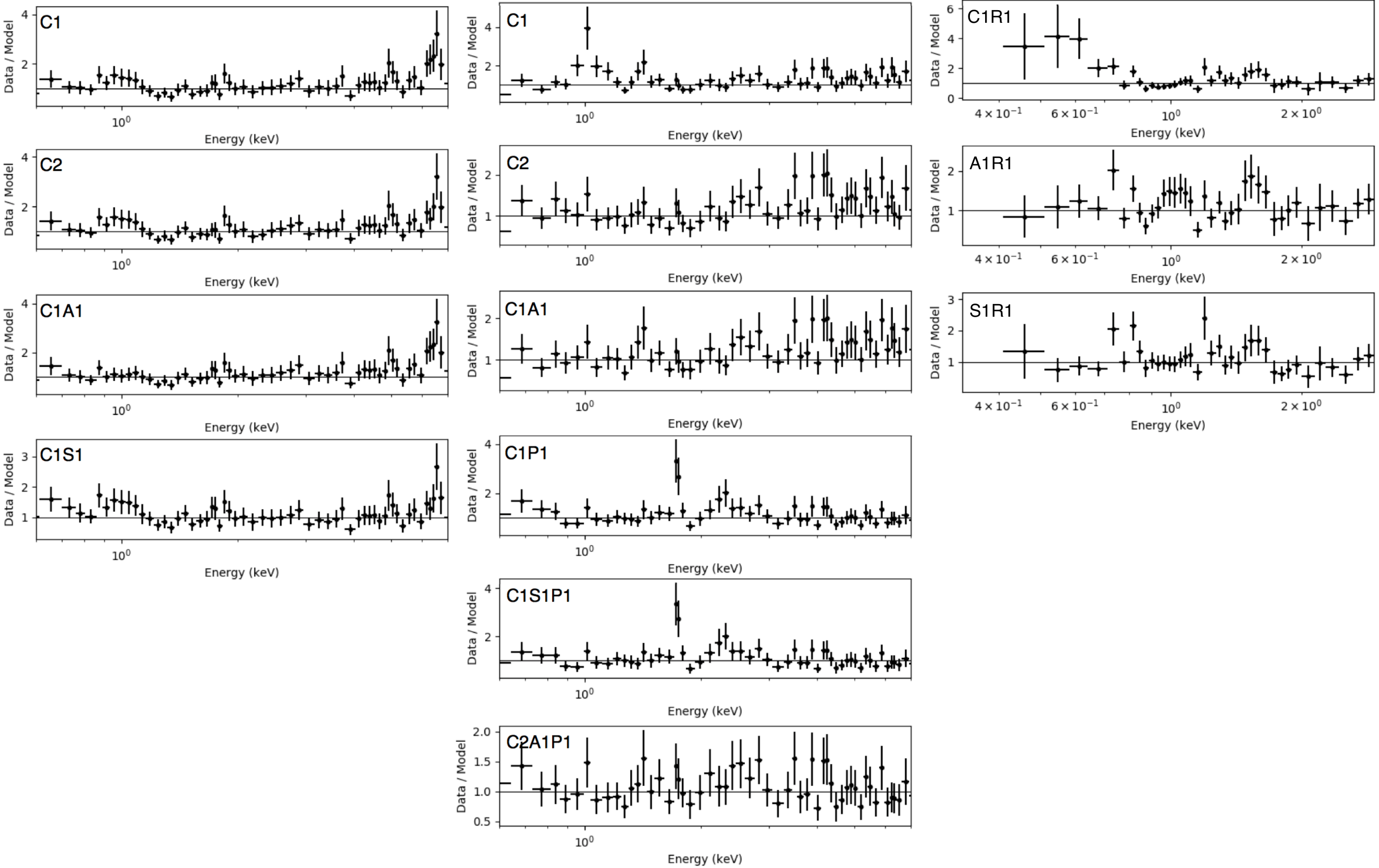}
   \caption{Residuals (data/model) obtained by fitting the NW (left), SE (center) cone and cross-cone (right) spectra with intermediate models used in the procedure described in Sect.~\ref{sect:ExtePhysicModels}. Models represent a combination of CLOUDY (C), APEC (A), PSHOCK (S), PEXMON (P) and PEXRAV (R) templates.}
   \label{image:ResCNW} 
   \end{center}
\end{figure}

\subsubsection{Cross-Cone} \label{sec:PhModCC}

The cross-cone spectrum (see Section~\ref{sec:extfitsph}) is different from that of the bi-cone. It shows an intense hard ($>$3 keV) X-ray continuum with a similar shape to the nuclear spectrum, and also a prominent soft excess around $\sim$1 keV, decreasing towards 2 keV. 
The hard spectral emission is dominated by spillover of the nucleus emission.
We fit the soft excess with photoionization, collisional and shock models.
We include a PEXRAV reflection component to fit the remaining $\sim 20 \%$ of the $>$3~keV emission, in excess of the nominal spillover predictions, as with the phenomenological model (Section~\ref{sec:extfitsph}). Most or all of the Fe K$\alpha$ line in the cross-cone spectrum ($\gsim$60$\%$) is due to the nuclear spillover. Because the physical models do not contribute to energies $\gsim$3~keV, we evaluate the fit-statistic only for the soft (0.3-3.0 keV) X-ray part of the spectrum. 

Adding model components to the baseline PEXRAV, we find a good statistics ($\chi_R^2 \gsim$1), but with wide contiguous residuals.
Using PEXRAV plus two-component models we obtain both a reduced $\chi_R^2 \approx 0.8$ and few residuals at $\sim$1.6 keV, at $<$2$\sigma$ significance. Adding additional components does not improve the fit. 

We show all the best-fit models in Fig.~\ref{image:CCPhMo} and the respective parameters in Table~\ref{table:parExtReg}.
The temperature of the gas in the collisional and shock phase is always less than or similar to $\sim$1.2 and 2.1~keV, respectively. The CLOUDY component implies the presence of photoionized gas with $log U \simeq 0.75$ and column density $log[N_H/cm^2] < 19.7$. The 2 CLOUDY components fit also requires gas with high photoionization parameter ($logU \simeq 1.75$).



\end{document}